\documentclass{aa}

\usepackage{graphicx}
\usepackage{upgreek}
\usepackage{makecell} 
\usepackage{longtable}
\usepackage{placeins}
\usepackage{txfonts} 

\usepackage{xtab,booktabs}
\usepackage{txfonts}
\usepackage{siunitx}
\usepackage{xcolor}
\usepackage[colorlinks,linkcolor=blue,citecolor=blue]{hyperref}
\usepackage{subfig}
\usepackage{mathtools} 
\usepackage{makecell} 
\usepackage{sidecap}

\addto\extrasenglish{
    
}
\addto\extrasenglish{
    
}
\addto\extrasenglish{
    
} 
\addto\extrasenglish{
    
}
\addto\extrasenglish{
    
}

\newcommand{\hi}{{\sc H\,i}}
\newcommand{\hii}{{\sc H\,ii}}

\let\upmu\muup

\def\aap{A\&A}

\def\apjs{ApJ\ Suppl.}
\def\apjl{ApJ\ Let.}

\begin{document} 

   \title{The longest known tails of ram-pressure-stripped star-forming galaxies are caused by an intracluster medium shock in Abell\,1367}
   \titlerunning{The longest known ram-pressure-stripped tails are caused by an ICM shock in A1367}

 \author{H. W. Edler
          \inst{1,2}
          \and
          M. Hoeft\inst{3}
          \and
          S. Bhagat\inst{3}
          \and
          A. Basu\inst{3,4}
          \and
          A. Drabent\inst{3}
          \and
          K. Rajpurohit\inst{5}
          \and
          M. Sun\inst{6}
          \and
          F. de Gasperin\inst{7}
          \and
          A. Botteon\inst{7}
          \and
          M. Br\"uggen \inst{2}
          \and
          A. Ignesti\inst{8}
          \and
          I. D. Roberts\inst{9,10}
          \and
          R. van Weeren\inst{11}
          }

       \institute{ASTRON, Netherlands Institute for Radio Astronomy, Oude
    Hoogeveensedijk 4, 7991 PD Dwingeloo, The Netherlands\\
                  \email{edler@astron.nl}
              \and 
              Hamburger Sternwarte, University of Hamburg,
              Gojenbergsweg 112, D-21029, Hamburg, Germany
              \and
              Th{\"u}ringer Landessternwarte, Sternwarte 5, D-07778 Tautenburg, Germany 
              \and
              Max-Planck-Institut f\"ur Radioastronomie, Auf dem H\"ugel 69, 53121 Bonn, Germany.
              \and
              Center for Astrophysics, Harvard \& Smithsonian, 60 Garden Street, Cambridge, MA 02138, USA
              \and 
              Department of Physics \& Astronomy, University of Alabama in Huntsville, 301 Sparkman Dr NW, Huntsville, AL 35899, USA
              \and
              INAF - Istituto di Radioastronomia,
              via P. Gobetti 101, Bologna, Italy
              \and
              INAF - Astronomical Observatory of Padova, vicolo dell'Osservatorio 5, IT-35122 Padova, Italy 
              \and
              Waterloo Centre for Astrophysics, University of Waterloo, 200 University Ave W, Waterloo, ON N2L 3G1, Canada
              \and
              Department of Physics \& Astronomy, University of Waterloo, Waterloo, ON N2L 3G1, Canada
              \and
              Leiden Observatory, Leiden University, PO Box 9513, 2300 RA Leiden, The Netherlands 
              }

   \date{Received November 6, 2025; accepted December 16, 2025}

  \abstract
   {The environment plays an important role in shaping the evolution of cluster galaxies through mechanisms such as ram pressure stripping (RPS), whose effect may be enhanced in merging clusters.}
   {We investigate a complex of three galaxies -- UGC\,6697, CGCG\,097-073, and CGCG\,097-079 -- that are currently undergoing extreme RPS, as is evident from their multiwavelength-detected tails. The galaxies are members of the nearby ($d=92$\,Mpc) merging cluster Abell\,1367 and are located in proximity to an intracluster medium (ICM) shock that is traced by X-ray observations and the presence of a radio relic.}
   {We analyzed LOFAR and MeerKAT observations at frequencies of 54, 144, 817, and 1270\,MHz to perform a detailed spectral analysis of the tails.}
   {We found that all three tails are significantly more extended than in previous radio studies, with lengths of $\geq70$\,kpc. For UGC\,6697, we detected a tail of $300$\,kpc, making it the longest known RPS tail of a star-forming galaxy at any wavelength. The length and spectral variations of the tail cannot be explained purely by the spectral aging of stripped cosmic rays. We constructed a model of the tail that includes compression and reacceleration due to the encounter with the nearby ICM shock, which can plausibly account for the extreme RPS as well as the length and spectral variation of the tail.
   We further discovered a radio plume at the leading edge of UGC\,6697 that connects to a narrow filament. These sources exhibit extremely steep ($\alpha\approx-1.7$) and highly curved spectra. We speculate that this emission arises from cosmic rays reenergized by 
   UGC\,6697’s rapid infall that propagate along magnetic filaments in the cluster center.}
   {Our findings represent direct evidence of a cluster merger shock impacting the evolution of member galaxies. Furthermore, we report the first tentative detection of  particle acceleration at the leading edge of an infalling galaxy.}
   
   \keywords{galaxies: clusters: individual: Abell 1367 -- radio continuum: general -- galaxies: interactions -- stars: formation}

   \maketitle
\nolinenumbers

\section{Introduction} \label{sec:intro}

Galaxy clusters form at the nodes of the cosmic web and are the most massive virialized structures in the Universe. The bulk of baryonic matter in galaxy clusters is confined in a hot ($T>10^7$\,K), low-density ($\rho \sim 10^{-27} \mathrm{g\,cm^{-3}}$) plasma known as the intracluster medium (ICM).
A smaller fraction of the baryons are in the form of stars, dust, and gas in the cluster galaxies, and cosmic rays from active galactic nuclei (AGNs). Galaxies that are members of a cluster are found to have systematically different physical properties compared to those inhabiting less dense cosmological environments. They show a reduced star formation rate (SFR; e.g., \citealt{Gavazzi1998TheDeterminations}) and lower neutral hydrogen gas contents (e.g., \citealt{Haynes1984TheGalaxies}) compared to galaxies of the same mass and morphological type outside of clusters.
These differences are caused by effects related to the environment, i.e., gravitational or hydrodynamical interactions that take place in clusters and affect the hydrogen content of the galaxies, which is the reservoir fueling star formation.
For massive clusters, the dominant mechanism responsible for quenching the SFR is thought to be ram pressure stripping \citep[RPS;][]{Gunn1972InfallMatterClusters,Boselli2022RamEnvironments}. In this process, the hydrodynamical drag forces due to the relative motion between galaxies and the ICM are sufficient to displace (part of) the interstellar medium (ISM) of infalling galaxies. The ram pressure scales as $P\propto\rho{v^2}$ and is thus strongest in massive clusters with high ICM densities, $\rho$, and for galaxies moving at a high velocity, $v$, relative to the ICM. Ongoing RPS events can be observed by the presence of one-sided asymmetric distributions of the neutral and/or ionized gas (traced by the \hi{} and H$\alpha$ lines, respectively; e.g., \citealt{scott2018Abell1367High,Poggianti2025MUSEViewRam}), the hot gas (traced in the X-rays, e.g., \citealt{Sun2022}), or the cosmic ray electrons (CRe; traced by the radio continuum, e.g., \citealt{Gavazzi1978WesterborkSurveyClusters}), which are all affected by the ram pressure. The stellar component in such galaxies is not significantly affected by the process because of their high mass and small cross section.

In the process of cluster formation, shock fronts are driven into the ICM, heating the thermal gas. Part of the shock energy is also dissipated into nonthermal components, namely relativistic particles and magnetic fields. Radio relics \citep[e.g.,][]{Jones2023PlanckClustersLOFAR} are diffuse sources colocated with cluster shocks where relativistic electrons are believed to be reaccelerated via diffusive shock acceleration (DSA, \citealt{Brunetti2014CosmicRaysGalaxy}), generating the synchrotron emission.
Due to ICM shocks and increased galaxy velocity dispersion, cluster mergers may increase the efficiency of RPS and thereby accelerate the evolution of galaxies in merging clusters. While it should, in general, lead to a quenching of the SFR, the compression of the ISM due to ram pressure may lead to a temporary phase of enhanced star formation \citep[e.g.,][]{Roberts2022LoTSSJellyfishGalaxiesa}.
\citet{Stroe2015RiseFallStar,Stroe2017LargeHaSurvey} found an enhancement of H$\alpha$ emitters in clusters with merger shock waves in the ICM, which they speculate traces enhanced star formation that is triggered by the merger activity. 
Conversely, \citet{Roberts2024shocks} found a statistically enhanced quenched fraction in clusters that host radio relics, which are a direct tracer of ICM shocks and thus cluster mergers. Similarly, \citet{Lourenco2023EffectClusterDynamical} found a mildly enhanced RPS fraction in interacting clusters. Further studies are required to understand the conditions under which and timescales in which the cluster dynamical state triggers or quenches the SFR of its member galaxies.

Abell\,1367 (hereafter A1367) is a relatively low-mass ($M_{200} = 3.8\times{10}^{14}\,M_\odot$) nearby ($d=92.2$\,Mpc) galaxy cluster with a virial radius of $r_{200}=1.18\,\mathrm{Mpc}$  \citep{Rines2003CAIRNSClusterInfall,Boselli2022RamEnvironments}. It is currently undergoing a major merger, as is evident from its elongated and bimodal ICM distribution \citep{Donnelly1998TemperatureStructureAbell}, non-Gaussian galaxy velocity distribution \citep{Cortese2004MultipleMergingAbell}, and the presence of a shock in the X-rays \citep{Ge2019MergerShockAbell} that corresponds to diffuse radio emission \citep{Gavazzi1978WesterborkSurveyClusters} that was classified as a radio relic by \citet{Ensslin1998ClusterRadioRelics,Farnsworth2013DISCOVERYMEGAPARSECSCALELOWa}.

In addition to the radio relic, the cluster hosts further complex and peculiar radio sources, likely connected to the ongoing merger. Among them is the radio galaxy 3C\,264 in the east of the cluster, which shows a prominent head-tail morphology \citep{Gavazzi1978WesterborkSurveyClusters}. 
The cluster also hosts a number of star-forming (SF) galaxies that are known to be undergoing RPS, as is evident by tails in the radio continuum and at other wavelengths. Three of them are located in the vicinity of the relic. The most spectacular case is the galaxy UGC\,6697, a starburst galaxy that was the first case of the detection of an RPS tail behind a SF galaxy \citep{Gavazzi1978WesterborkSurveyClusters,Gavazzi198750KPCRadio}. The extent of the tail is reported to be $\sim30$\,kpc at 1.5\,GHz \citep{Gavazzi1995peculariA1367}. It is coincident with a 120\,kpc long tail of hot gas traced in the X-rays  \citep{Sun2022}, a tail of ionized gas that shows an extent of 100\,kpc in MUSE observations \citep{Consolandi2017UGC6697} and an \hi{}-tail of at least 70\,kpc \citep[e.g.,][]{Scott2010ProbingEvolutionaryMechanisms}.
The other two galaxies in the relic region, CGCG\,097-079 and CGCG\,097-073, hereafter referred to as C\,079 and C\,073,  also show remarkably long radio and H$\alpha$ tails \citep{Gavazzi1978WesterborkSurveyClusters,Gavazzi1995peculariA1367,Roberts2021LoTSSJellyfishGalaxies}. 

The complex including the three tailed galaxies and the radio relic in A1367 is a tentative example of a direct interaction between a large-scale ICM shock and RPS of SF galaxies. The three highly peculiar SF galaxies show long multiwavelength tails with a similar orientation (their position angles are within $40^\circ$). Their tails are mostly in the propagation direction of the shock and they are starburst galaxies, with strong star formation taking place in particular at their leading edge \citep{Consolandi2017UGC6697,Pedrini2022MUSESneaksPeek}. This makes the scenario of an interaction highly plausible.

In this paper, we present an analysis of LOw-Frequency ARray (LOFAR, \citealt{vanHaarlem2013LOFARLOwFrequencyARray}) observations of the A1367. The high sensitivity of this telescope at low and ultralow frequencies allows us to be sensitive to the steep-spectrum emission that is expected in stripped tails \citep[e.g.,][]{Roberts2024RadiocontinuumSpectraRampressurestripped}. We further present the analysis of MeerKAT \citep{Jonas2009MeerKATSouthAfrican} data, allowing us to constrain radio spectra in A1367 accurately.

In this work, we assume a flat $\mathrm{{\Lambda}}$ cold dark matter ($\mathrm{{\Lambda}CDM}$) cosmology with $\Omega_\mathrm{m}=0.3$ and $H_0=70\,\mathrm{km\,s^{-1}\,Mpc^{-1}}$. At the distance of the cluster, 1\,arcmin corresponds to 27\,kpc. 
This paper is arranged as follows: in \autoref{sec:data}, we list the radio observations, data reduction, and image analysis, in \autoref{sec:radioemission}, we discuss the emission of the three galaxies individually, and in \autoref{sec:tailanalysis}, we present a detailed analysis and modeling of the UGC\,6697 tail. The radio filaments are covered in \autoref{sec:filament}, while we discuss our findings in \autoref{sec:discussion}. We conclude in \autoref{sec:conclusion}.

\section{Data reduction and analysis}\label{sec:data}

\begin{table}
\centering\small
\caption{Observations. }\label{tab:metadata}
\begin{tabular}{ l c c c c }\hline
Freq. band & Obs. date & Time & Freq. & Bandwidth \\ 
 &  & [$h$] & $[$MHz$]$ &  $[$MHz$]$ \\ \hline\hline\vspace{1mm}
LBA & 02 Jan. 2020 & 6 & 54 & 24 \\
 & 13 March 2020 & 6 & 54 & 24 \\\hline
HBA &  04 Nov. 2019 & 8 & 144 & 48\\
  &  13 March 2020 & 8 & 144 & 48\\
  &  10 Oct  2020 & 8 & 144 & 48\\
  &  11 Oct 2020 & 8 & 144 & 48\\\hline
UHF-band &  25 Feb 2024 & 4 & 817 & 410\\\hline
L-band &  06 June 2020 & 3.3 & 1270 & 778\\
 &  10 June 2020 & 3.3 & 1270 & 778\\
 &  13 June 2020 & 3.3 & 1270 & 778\\
\end{tabular}
\label{tab:obs}
\end{table}

In this paper, we analyze a multifrequency radio dataset. We use radio observations with the LOFAR low-band antenna (LBA; 42--68\,MHz), LOFAR high-band antenna (HBA; 120--168\,MHz), MeerKAT ultra-high frequency (UHF; 580--1015\,MHz), and MeerKAT L-band (900--1670\,MHz).
Details about the observations can be found in \autoref{tab:obs}. 

\subsection{LOFAR LBA}
Two observations of 6\,h each were conducted on January 2 and March 17, 2020, in the frequency range 42--68\,MHz. As a primary calibrator source, 3C\,295 was observed during the whole observation in parallel.
Processing was carried out using the \texttt{Library for Low-frequencies}\footnote{\url{https://github.com/revoltek/LiLF}} (LiLF), the standard calibration framework for LOFAR LBA. Initially, the data were downloaded and averaged to 4\,s time and 49\,kHz frequency resolution. 
The instrumental systematic effects were calibrated on 3C\,295, which was observed during the whole observation with a parallel beam, and applied to target field dataset.
Then, we proceeded with the direction-independent self-calibration of the target field as described in \citet{deGasperin2020ReachingThermalNoise}, with one modification: the original strategy derives direction-independent calibration solutions for the full target field in two self-calibration iterations. We instead carried out the second self-calibration cycle only on the central bright source 3C\,264, after subtracting the rest of the field with the solutions found in the first cycle. 3C\,264 is sufficiently bright to obtain robust calibration solutions and is part of the A1367 cluster, so we optimize solutions for the region of the field in which we are interested. 
After correcting the data for the ionospheric solutions toward 3C\,264, we carry out wide-field direction dependent calibration according to the strategy described in \citet{Edler2022Abell1033Radio}.
In this way, direction-dependent calibration solutions are obtained across the field of view (FoV). These are then applied on-the-fly during imaging to individual facets. We use the facet-imaging mode of the \texttt{WSClean} imager \citep{Offringa2014WSCLEANImplementationFast,Offringa2017OptimizedAlgorithmMultiscale} to create the final wide-field corrected image. 

For the A1367 region, this image still shows significant calibration artifacts due to the bright  radio galaxy 3C\,264. Thus, we employ the scheme of source extraction and re-calibration, originally described in \citet{vanWeeren2021LOFARObservationsGalaxy}. For this, we subtract all sources in the FoV except for our region of interest, a $1^\circ\times1^\circ$ square region centered on the A1367 cluster. We then perform iterations of direction-independent self-calibration of this area. We solve for scalar-phases on time intervals of 8\,s, constraining the solutions to be smooth on scales of a 3\,MHz Gaussian kernel. After five cycles, we perform a final self-calibration cycle where we additionally solve for scalar amplitudes on a 8\,s time interval and a full-Jones matrix on 128\,s time intervals. This procedure strongly mitigates the artifacts originating from 3C\,264. At this point, the dominant systematic effects in the data are direction-dependent ionospheric effects within the extraction region, which is larger than the coherency scale of the ionosphere. Thus, after applying the solutions found for the A1367 region to the data, we perform three more rounds of direction-dependent scalar-phase calibration, using 3C\,264 and three other bright sources in the field as direction-dependent calibrators.

\subsection{LOFAR HBA}
Four observations of 8\,h each were considered for this work, three of which target the cluster center (project LC13\_024), the fourth one is centered on 3C\,264 (LC11\_016). Initial calibration followed the standard LOFAR HBA calibration routine, using the pipelines \texttt{LINC} \citep{deGasperin2019SystematicEffectsLOFAR} and \texttt{ddf-pipeline} \citep{Tasse2021LOFARTwoMeter}.
Then, we again carried out the extraction and re-calibration procedure described above using a $0.9^\circ$ square region of interest. In this step, sources outside the region of interest were subtracted and the measurement sets were phase-shifted to the cluster center. Then, it was solved for scalar phases at 16\,s time resolution. After three iterations, we additionally carried out a scalar amplitude solve on a 3\,min timescale. In both of these solves, a frequency resolution of 0.39\,MHz was used, but the effective degrees of freedom were reduced by enforcing spectral smoothness using a Gaussian kernel of 5\,MHz width. Finally, after six iterations, we also solved for a full-Jones matrix every 11\,min and 1.95\,MHz for three more iterations.
Then, after applying the solutions of the direction-independent extraction procedure, we again carried out direction-dependent calibration toward the four brightest sources in the region of interest and solving for scalar phases on a time-resolution of 64\,s and enforcing spectral smoothness with an 8\,MHz kernel.

\subsection{MeerKAT} 
MeerKAT observations were taken as part of projects SCI-20210212-MH-01 (L-band; 
PI M. Hoeft) and SCI-20230907-MH-01 (UHF-band; PI M. Hoeft). 
The observations were calibrated following standard procedures, the details of the data reduction will be described in an accompanying paper (Hoeft et al. in prep.).
The final MeerKAT images were corrected for the primary beam effect using \texttt{katbeam} \footnote{\url{github.com/ska-sa/katbeam}}.

\subsection{Image analysis}\label{sec:imaging}
\begin{figure}
    \centering
    \includegraphics[width=0.99\linewidth]{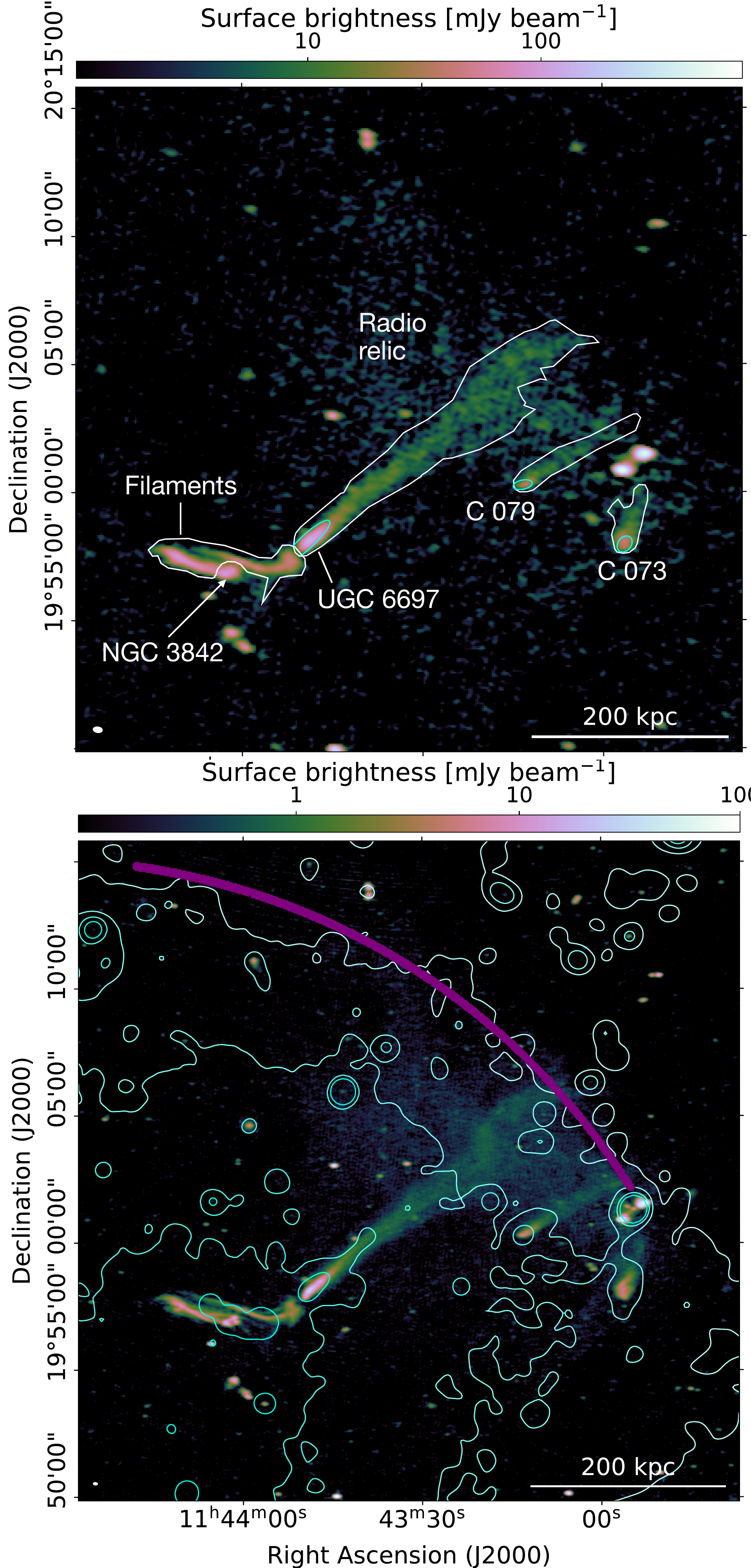}
    \caption{Top panel: Northwest of A1367 at 54\,MHz, radio sources discussed in the text are highlighted. White and cyan regions were used to measure the total and disk flux densities, respectively. 
    Bottom panel: Same region at 144\,MHz, also displaying the XMM-Newton X-ray contours of \citet{Ge2019MergerShockAbell}, contours are $[5,10,20,40]\,\mathrm{counts\,s^{-1}\,deg^{-2}}$. The purple region marks the detected ICM shock. 
    }
    \label{fig:wide}
\end{figure}

All radio maps in this work were imaged using \texttt{WSClean} and were corrected for the primary beam attenuation. The final LOFAR maps, created with a robust weighting parameter of $-0.5$, are shown in \autoref{fig:wide}.
We investigated the astrometry of our images to ensure there are no offsets which affect our analysis. For this, we employed \texttt{PyBDSF} \citep{Mohan2015PyBDSFPythonBlob} to create source catalogs at each frequency. Using only point sources and with the HBA map as reference, we determined a flux-weighted offset of $\Delta\mathrm{RA}=-2.8''$; $\Delta\mathrm{Dec}=3.3''$ for the LOFAR LBA, $\Delta\mathrm{RA}=-0.3''$; $\Delta\mathrm{Dec}=0.1''$ for the MeerKAT UHF and $\Delta\mathrm{RA}=-0.9''$; $\Delta\mathrm{Dec}=-0.6''$ for the MeerKAT L-band image. For the LBA, the offset is approximately a third of the point-spread function (PSF), while for the MeerKAT, it is below a tenth of the PSF. We shifted the images to correct for this offset.

In \autoref{fig:wide}, we mark the radio sources that are the subjects of this work. The elliptical galaxy NGC\,3842 and the connected filaments are a secondary subject of our study and will be discussed in \autoref{sec:filament}. The three SF galaxies UGC\,6697, C\,079, and C\,073 and their continuum tails are clearly visible.  The diffuse emission of the radio relic is partly superimposed on the emission of the tails. 
Based on the LOFAR HBA map at $21''$ resolution, we manually defined the regions corresponding to the three tailed galaxies by outlining the elongated area of high surface brightness behind the galaxies. These manual regions are shown as white contours in \autoref{fig:wide}. 

\subsubsection{Subtraction of the radio relic emission}
For the analysis of the stripped tails, the co-spatial diffuse emission of the radio relic could impose biases. Thus, we considered two different scenarios: first, we did not subtract the diffuse relic emission from the tail, assuming that it can be neglected in comparison to the tail emission. Secondly, we attempted to model and subtract the diffuse relic emission from the tails, assuming that it is fully independent of those and can be reasonably interpolated from the emission surrounding them.
To model the emission of the radio relic, we first created images based on visibilities from which compact background sources were subtracted to allow for an accurate interpolation of the diffuse emission.
For this, we created  a model of the compact background sources in our region of interest by imaging the data at high angular resolution using a robust weighting of $-1.0$ and an inner $uv$ cut of 2000$\lambda$ while applying a clean-mask containing the background sources. 
We then predicted the visibilities corresponding to the compact source model, corrupted them with the direction-dependent calibration solutions (for LOFAR data), subtracted those from the data, and imaged the subtracted data. 
\begin{figure}
    \centering
    \includegraphics[width=0.6\linewidth]{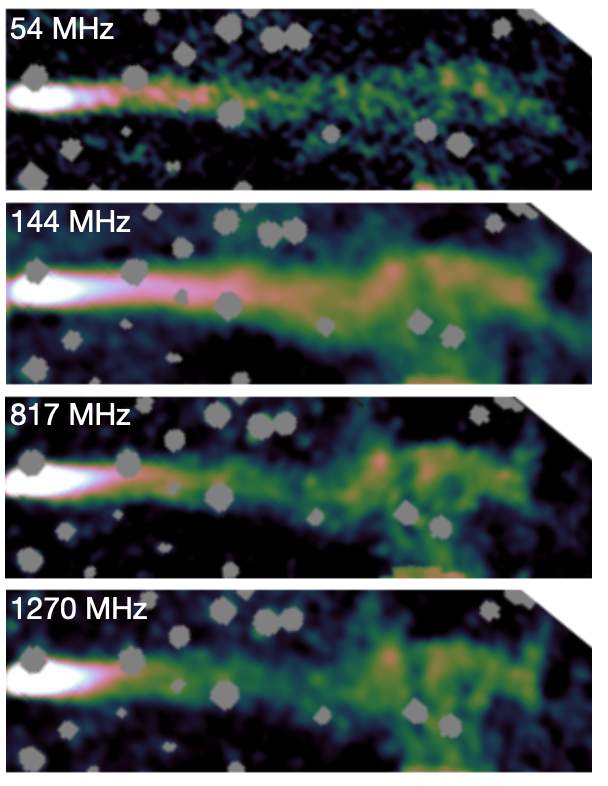}
    \caption{UGC\,6697 at $21''$ resolution after subtracting the radio relic model. Gray areas show masked sources. For display purposes, the images are rotated by 40$^\circ$.}
    \label{fig:relicsub_4}
\end{figure}

\begin{figure}
    \centering
    \includegraphics[width=1.0\linewidth]{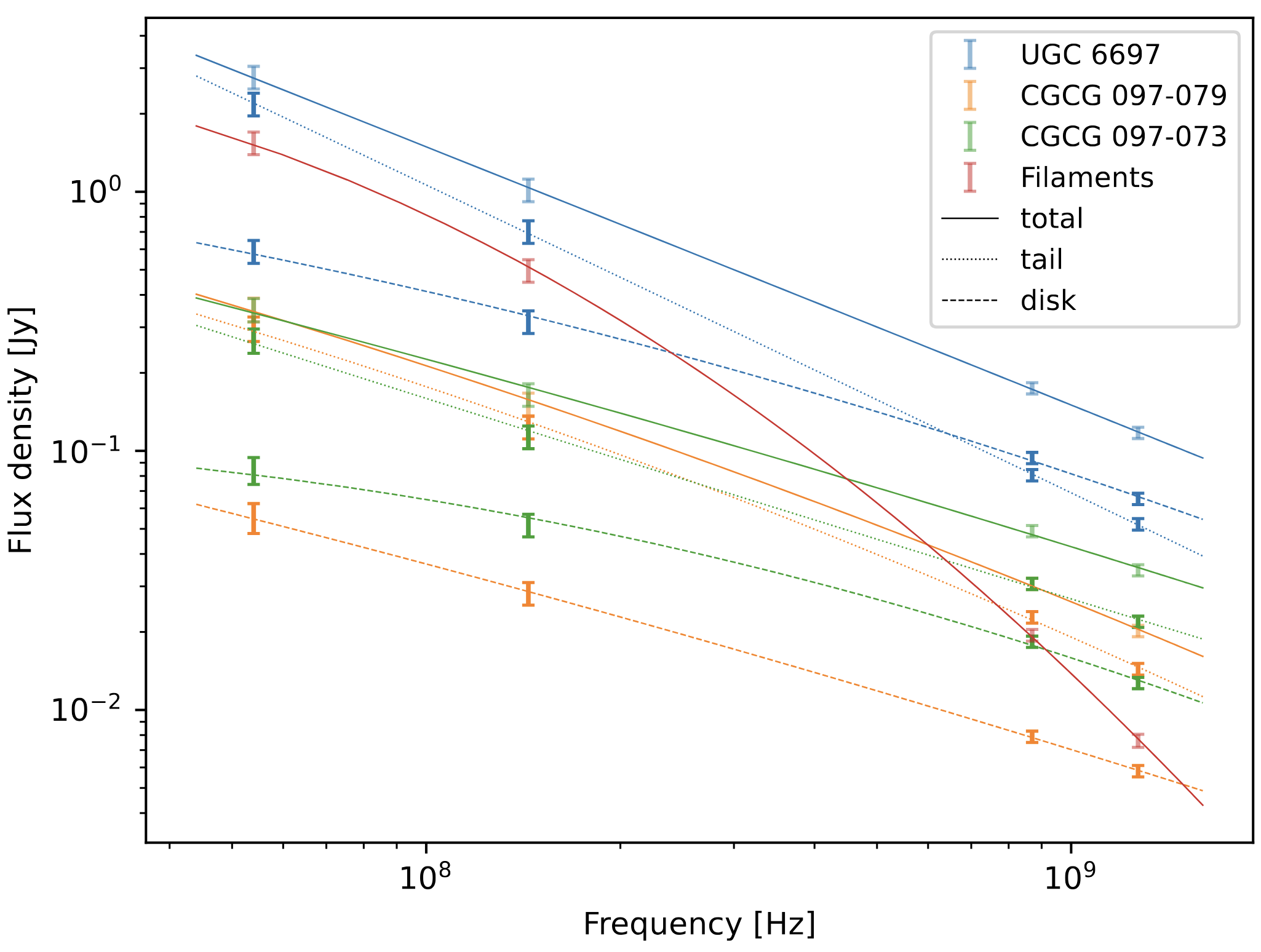}
    \caption{Radio spectrum of UGC\,6697 (blue), C\,079 (orange), and C\,073 (green). The lines are log-parabolic fits to the data, where the solid lines were fit to the total flux density, and the dotted and dashed ones to the flux in the tail and disk, respectively. In red, we also show the filaments close to NGC\,3842 discussed in \autoref{sec:filament}.}
    \label{fig:fluxes}
\end{figure}

Based on these compact-source subtracted maps, we modeled the emission of the relic, using rectangular regions which are $40''\times350''$ wide ($40''\times200''$ for C\,079) and spaced evenly along the tail, perpendicular to it. For UGC\,6697, we display these regions in \autoref{sec:appendix_relicmodel}. For each region, we calculate the mean surface brightness, excluding the area that overlaps with the tail. This interpolation of the relic surface brightness is then used to create a model of the diffuse background emission. It is then convolved with a Gaussian kernel ($\sigma_\mathrm{smooth}=20''$) to obtain a smooth model. As an example, one such model is shown in  \autoref{sec:appendix_relicmodel}. This model is then subtracted from the image, this procedure is carried out for all four frequencies. For C\,073, it is less clear which part of the emission is related to the tail or the radio relic. Since the radio relic emission is anyway significantly fainter at the location of this galaxy, no background subtraction was attempted. 
In \autoref{fig:relicsub_4}, we display the UGC\,6697 tail after subtracting the relic emission at all four frequencies. The tail remains as a well-defined morphological feature.

\subsubsection{Flux density and spectral index measurements}

\begin{table}
\centering
\small
\caption{Spectral fit results. }
\begin{tabular}{lccc}
\hline\hline
 & $S_{300}$ [Jy] & $\alpha$ & $\beta$ \\
\midrule
U6697 & $0.500\pm0.044$ & $-1.00\pm0.07$ & $-0.00\pm0.25$ \\
U6697 disk & $0.205\pm0.018$ & $-0.70\pm0.07$ & $-0.13\pm0.25$ \\
U6697 tail & $0.289\pm0.026$ & $-1.19\pm0.07$ & $-0.01\pm0.25$ \\
C079 & $0.083\pm0.007$ & $-0.91\pm0.07$ & $-0.10\pm0.25$ \\
C079 disk & $0.017\pm0.002$ & $-0.71\pm0.08$ & $-0.05\pm0.27$ \\
C079 tail & $0.066\pm0.006$ & $-0.96\pm0.07$ & $-0.13\pm0.25$ \\
C073 & $0.105\pm0.009$ & $-0.72\pm0.07$ & $-0.05\pm0.25$ \\
C073 disk & $0.037\pm0.003$ & $-0.60\pm0.08$ & $-0.20\pm0.26$ \\
C073 tail & $0.068\pm0.006$ & $-0.77\pm0.07$ & $0.01\pm0.25$ \\
Filament & $0.164\pm0.015$ & $-1.74\pm0.07$ & $-0.60\pm0.26$ \\
\bottomrule
\end{tabular}
\tablefoot{Column $S_{300}$ refers to the fit flux density at 300\,MHz, and $\alpha$ and $\beta$ are the spectral index and curvature as defined in \autoref{eq:logparabola}.}
\label{tab:spectralfit}
\end{table}
To measure the integrated flux densities of the RPS galaxies, we used the source regions defined previously and displayed in \autoref{fig:wide}. The values were measured using the same source regions at all frequencies and using images convolved to a circular PSF of $21''$. For measurements of the emission from the stellar disks, we defined the disk regions using the $r$-band isophotal diameter at a reference level of $25\,\mathrm{mag\,arcsec^{-2}}$ \citep{SDSS2007DR5}. The emission from the disks was measured using the high-resolution maps with a PSF of $14''\times10''$ to allow for a better separation of the radio emission in the disk and the tail.
The flux density of the tails was calculated as the difference between the total flux density of the galaxies and the flux density of the disk.
We estimated integrated spectral index values $\alpha$ and spectral curvature values $\beta$ by fitting a second-order polynomial in log$_{10}$-log$_{10}$-space (hereafter log-parabola) to the four observed frequencies according to:
\begin{equation}\label{eq:logparabola}
    \log{S(\nu)} = \log{S_0} + \alpha\log{\left(\frac{\nu}{\nu_0}\right)} + \beta\log^2{\left(\frac{\nu}{\nu_0}\right)};
\end{equation}
where $\nu_0=300\,$MHz.
We assumed a systematic flux density scale uncertainty of 10\% for the LBA and the HBA maps \citep{Shimwell2022LOFARTwometreSky,deGasperin2023LOFARLBASky}, and a 5\% uncertainty for the flux density calibration in MeerKAT \citep{Knowles2022MeerKATGalaxyCluster}. 
The best-fitting values are listed in \autoref{tab:spectralfit}. We plot the log-parabola fits together with the measurements in \autoref{fig:fluxes}. 
For all three galaxies, we find integrated spectra with spectral indices (including the tail emission) between $-0.72\pm0.07$ and $\alpha=-1.00\pm0.07$, steeper than normal SF galaxies in the nearby Universe ($\langle\alpha_{144}^{1400}\rangle\approx -0.56\pm0.14$, \citealt{Heesen2022NearbyGalaxiesLOFAR}). For all three galaxies, the tail spectra are steeper than the disks, in agreement with the findings of \citet{Ignesti2023RadioContinuumTails,Roberts2024RadiocontinuumSpectraRampressurestripped}. The spectra of the disks alone are within the scatter of the spectral indices of SF galaxies. 
The spectra of the galaxies do not significantly deviate from a power-law spectrum; however, the high uncertainty on $\beta$ means we cannot exclude the presence of moderate spectral curvature.

\begin{figure}
    \centering
    \includegraphics[width=0.55\linewidth]{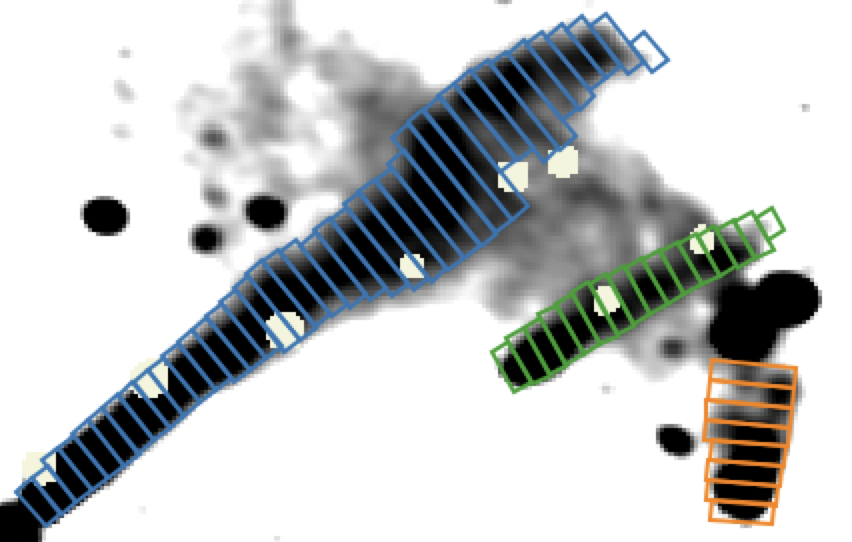}
    \caption{Regions used to measure the flux density along the tails, spaced at $21''$, identical to the resolution of the background map at 144\,MHz. Beige regions are masked background sources.}     \label{fig:tailregs}
\end{figure}

\begin{figure}
    \centering
    \includegraphics[width=0.99\linewidth]{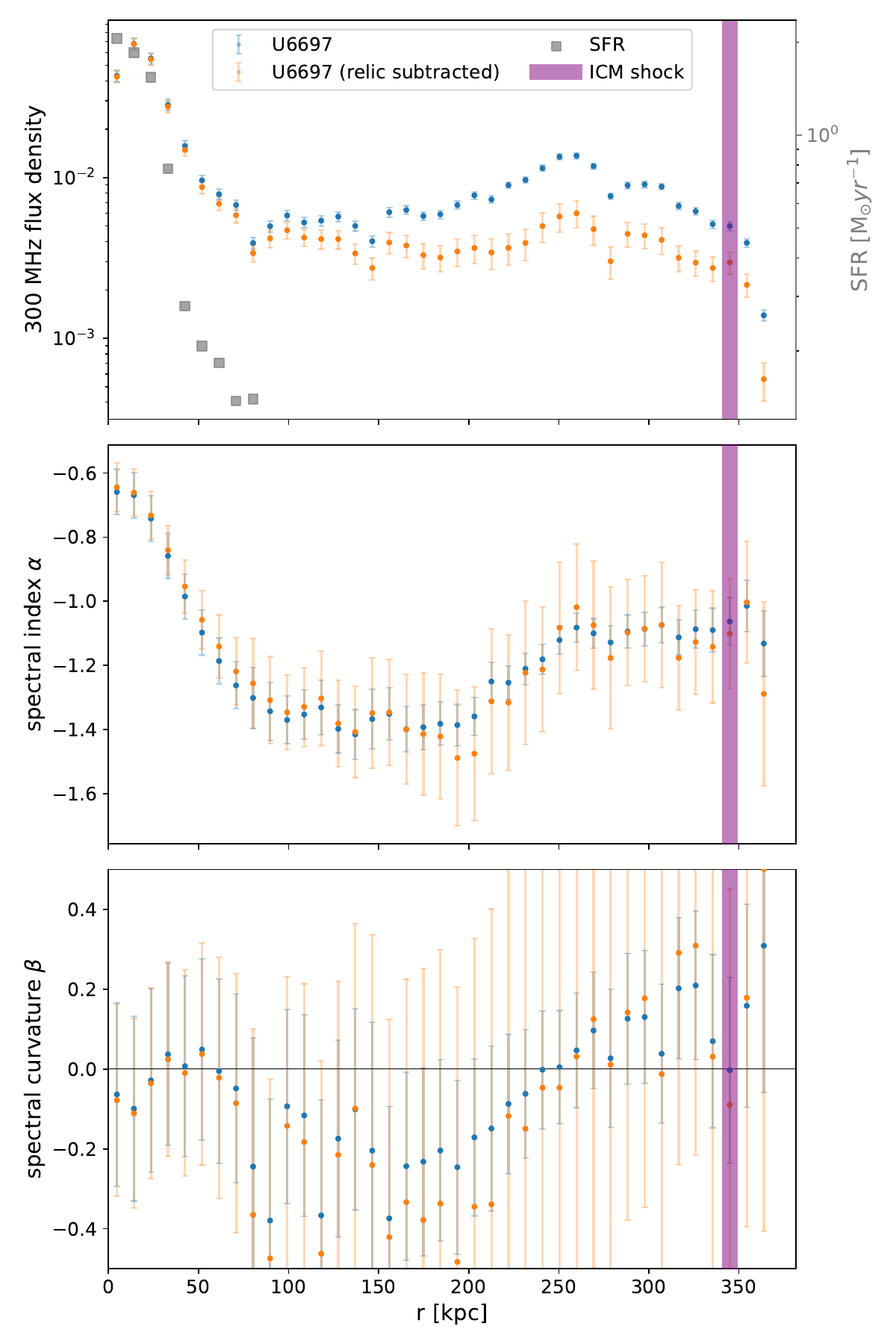}
    \caption{Flux density (top panel), SFR (top panel, right $y$ axis), spectral index (mid panel), and spectral curvature (bottom panel) as a function of projected distance along the UGC\,6697 tail. The vertical purple line shows the location of the ICM shock.}
    \label{fig:si_tails}
\end{figure}
We also measured the flux density trend along the tails from the $21''$ resolution images. We decomposed the regions covering the tails in segments spaced by one PSF, corresponding to a projected distance of 9.45\,kpc, as displayed in \autoref{fig:tailregs}. For each of these regions, we measured the flux densities at the four frequencies and fit \autoref{eq:logparabola} to the measured values.
To ensure that spectral trends along the tail are not driven by contamination from the radio relic, we repeated this procedure using the radio-relic-subtracted maps. The resulting plots for UGC\,6697 are shown in \autoref{fig:si_tails}; for C\,079 and C\,073, they are displayed in \autoref{sec:appendix_relicmodel}.

\subsubsection{Spectral index maps}
\begin{figure}
    \centering
    \includegraphics[width=0.99\linewidth]{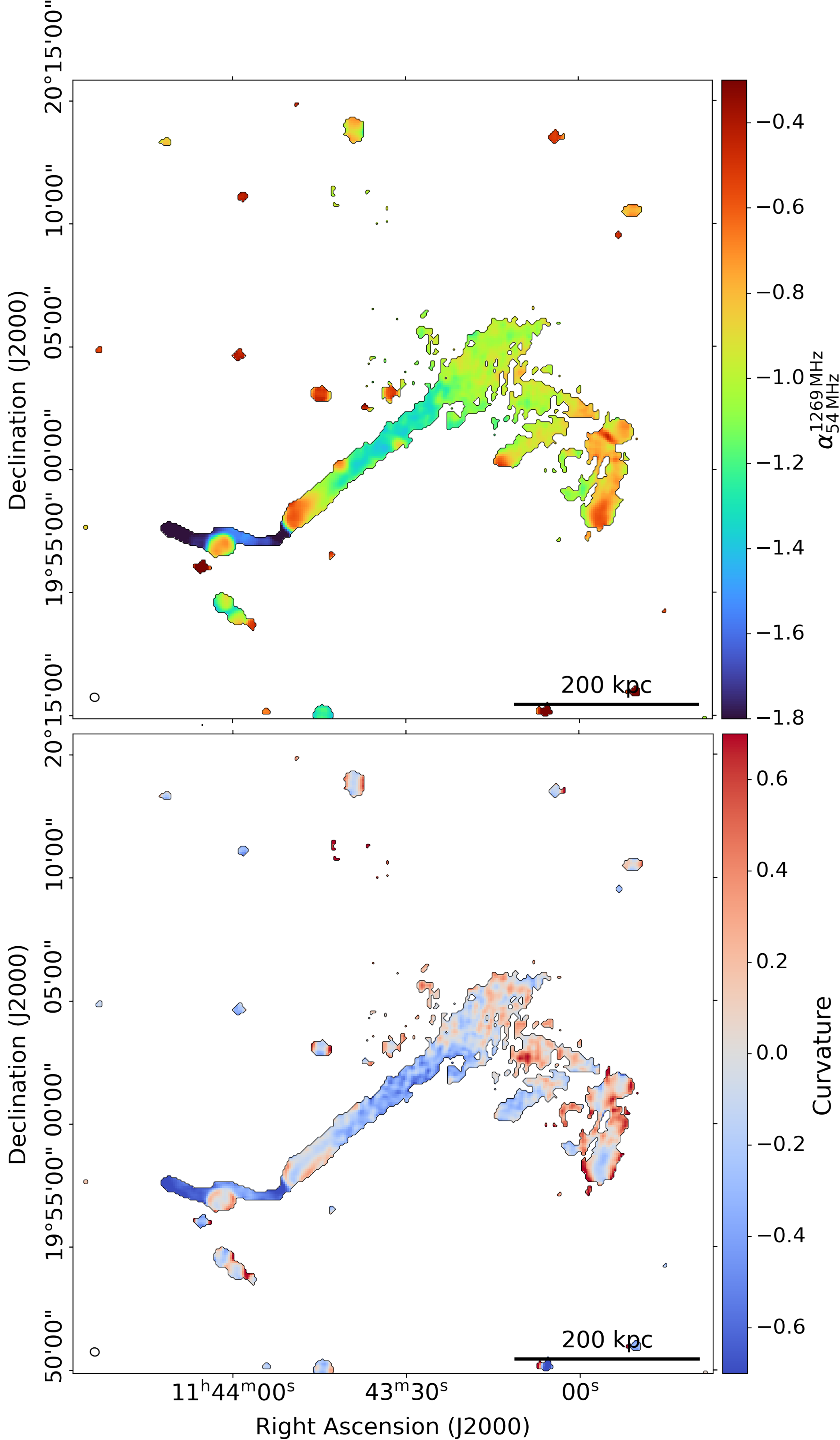}
    \caption{Top panel: Map of the spectral index, $\alpha$. Bottom panel: Map of the spectral curvature, $\beta$. 
    The angular resolution of the maps is $21''\times20''$.}
    \label{fig:si}
\end{figure}
To create spectral index maps, we started from the radio maps with a harmonized $uv$ coverage by applying an inner cut of 150$\lambda$. Then, we convolved the maps at all frequencies to a common resolution of $21''\times20''$.  Next, we re-gridded the images to align them on the same pixel grid. For pixels that are above $3\sigma_\mathrm{rms}$ in all images, we fit  \autoref{eq:logparabola} to retrieve the flux density at $\nu_0=300\,$MHz, the spectral index, the spectral curvature, and the corresponding uncertainties.  
The resulting spectral parameter maps are displayed in \autoref{fig:si}, the corresponding uncertainty maps can be found in \autoref{sec:appendix_spec}.

\section{The radio emission of the stripped galaxies}\label{sec:radioemission}
Our LOFAR images in \autoref{fig:wide} reveal that the radio tails behind the three stripped SF galaxies UGC\,6697, C\,079, and C\,073 in the west of the cluster are all significantly more extended than reported in previous studies using higher frequency and/or lower sensitivity radio data \citep[e.g.,][]{Gavazzi198750KPCRadio,Gavazzi1995peculariA1367,Farnsworth2013DISCOVERYMEGAPARSECSCALELOWa,Roberts2021LoTSSJellyfishGalaxies}.
Here, we compare their radio emission to other SFR tracers, followed by a discussion of the radio morphology, spectra, and the general properties of the three objects. 

\subsection{Radio-SFR relation}
\begin{figure}
    \centering
    \includegraphics[width=0.99\linewidth]{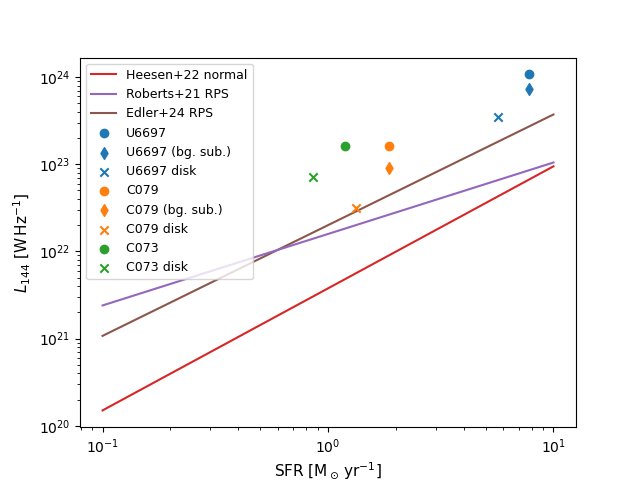}
    \caption{Luminosity at 144\,MHz ($y$ axis) compared to FUV-MIR derived SFR ($x$ axis). 
    The lines show radio-SFR relations from the literature, \citet{Heesen2022NearbyGalaxiesLOFAR} (red) is derived for normal (non-RPS) galaxies using total infrared as tracer, \citet{Roberts2021LoTSSJellyfishGalaxies} and \citet{Edler2024ViCTORIAProjectLOFAR} are derived for RPS galaxies using spectral energy density fitting and FUV+MIR, respectively.}
    \label{fig:radiosfr}
\end{figure}

We calculated the total SFR for the galaxies using the far-UV (FUV) emission, which we dust absorption-corrected with the mid-infrared (MIR). The SFR of the galaxies likely had significant recent variation given their infall onto A1367. The FUV emission is primarily due to stars formed $\sim100$\,Myr ago,
similar to the lifetime of the synchrotron emitting CRes below 1\,GHz \citep{Hao2011DustcorrectedStarFormation}. Thus, FUV and 144\,MHz continuum trace SFR on comparable timescales.
We measured the FUV luminosities $L_\mathrm{FUV}$ from GALEX images \citep{Bai2015,Bianchi2017}. These were corrected for dust extinction using the  $22\,\mu{m}$ WISE band 4 images \citep{WISE2012}. The SFR was calculated for the initial mass function of \citet{Kroupa2001VariationInitialMass} following \citet{Hao2011DustcorrectedStarFormation}.
The galaxies have a total SFR of 7.8, 1.9, and $1.2\,\mathrm{M_\odot\,yr^{-1}}$ for UGC\,6697, C\,079, and C\,073, respectively. 
Within the (old) stellar disk, as defined by the $r$-band isophotes, the SFR is 5.7, 1.3, and $0.9\,\mathrm{M_\odot\,yr^{-1}}$. In \autoref{fig:si_tails}, we display the SFR in 9.5\,kpc spaced regions along the disk and tail of UGC\,6697. Star formation is also taking place outside of the disk along the initial part of the tail. For the other two galaxies, star formation takes place only in the direct vicinity of the stellar disk and does not extend into the tails for more than 10\,kpc.

In \autoref{fig:radiosfr}, we plot the 144\,MHz luminosity against the SFR$_\mathrm{FUV+22\mu{m}}$ and compare to relations from the literature. The total continuum luminosity of the three galaxies is a factor $\sim$ 20 -- 30 higher compared to relations found for normal galaxies in the nearby Universe \citep{Heesen2022NearbyGalaxiesLOFAR}. Similarly, \citet{Gavazzi1995peculariA1367} found the three galaxies to be a factor of ten more radio luminous than expected given their infrared emission, which also traces the SFR. While RPS galaxies are known to have excess radio emission compared to normal galaxies \citep{Chen2020RamPressureStripped}, even compared to relations for RPS galaxies of \citet{Roberts2021LoTSSJellyfishGalaxies,Edler2024ViCTORIAProjectLOFAR}, all three galaxies are still a factor of 5-6 more radio luminous. This reduces to an offset of $\sim3$ when considering the relic-subtracted radio maps described in \autoref{sec:imaging}. Considering only the stellar disk, we still find an  excess compared to other RPS galaxies that ranges from 1.6 for C\,079 to 5.1 for C\,073.   
Thus, the excess radio emission does not just originate from the tails. Our results further suggest a trend in which  objects which are experiencing more extreme RPS (such as the three galaxies considered here) show a stronger radio excess compared to the general population of RPS galaxies \citep[e.g.,][]{Roberts2021LoTSSJellyfishGalaxies,Edler2024ViCTORIAProjectLOFAR}. 
A plausible explanation of the enhanced radio emission is a higher magnetic field strength due to the ISM turbulence and compression caused by the RPS, furthermore, variable SFR can play a role \citep{Ignesti2022WalkLowSide, Edler2024ViCTORIAProjectLOFAR}.

\subsection{UGC 6697}
UGC\,6697 is a large ($M_\mathrm{star}=2\times10^{10}\,\mathrm{M}_\odot$) irregular starburst galaxy known to host multiwavelength tails \citep{Consolandi2017UGC6697,Sun2005RevealingInteractionXRay}. 
Our LOFAR images reveal that the total extent of UGC\,6697 is 365\,kpc ($14.6'$), the final 310\,kpc of which are the emission of the tail. This means that the radio tail is considerably more extended than the $\approx30$\,kpc found in previous works \citep{Gavazzi1995peculariA1367}.  This makes UGC\,6697 by far the SF galaxy with the longest known tail, more than a factor of three longer than the largest previously known radio tail (NGC\,2276 with 100\,kpc, \citealt{Roberts2024NGC2276}) and also surpassing the current record holder across all wavelengths, NGC\,4569's tail of 230\,kpc in the H$\alpha$ \citep{Sun2025VirgoEnvironmentalSurvey}. While the tail is partly superimposed on the diffuse relic emission, the radio morphology clearly suggests that it is a well-defined feature and not just blending in with the general relic emission. This is particularly evident in the maps we obtain after subtracting a model of the relic emission in \autoref{fig:relicsub_4}.
In addition, as already noted, UGC\,6697 is also extremely radio-luminous compared to its SFR, being a factor $\approx{13}$ above the radio-SFR relation for normal nearby SF galaxies of \citet{Heesen2022NearbyGalaxiesLOFAR} (see orange circle and red line in \autoref{fig:radiosfr}). 

In the central panel of \autoref{fig:si_tails}, we show the spectral index trend along the tail for UGC\,6697 as a function of the projected distance. In the disk, the spectral index is $
\alpha=-0.70\pm0.07$, in line with the expectation for SF galaxies. Until $r\approx80\,$kpc, the spectrum quickly steepens toward a value $\alpha \approx -1.4$. Afterward, it stays approximately constant until $r\approx200\,$kpc, after which the spectrum begins to gradually flatten to $\alpha \approx -1.1$. The reason for this spectral behavior will be investigated in \autoref{sec:tailanalysis}.

\subsection{CGCG 097-079}
In this galaxy, \citet{Scott2015HighlyPerturbedMolecular} found that the ISM is perturbed even in the molecular phase, which is more deeply embedded in the gravitational potential well. They attributed this to the ram pressure, possibly in combination with a past tidal interaction with C\,073. They find evidence for a burst of SF $\sim100$\,Myr ago. 
The galaxy also hosts long tails in the H$\alpha$ \citep{Boselli2014OriginFaintendRed,Pedrini2022MUSESneaksPeek}, X-ray \citep{Sun2022}, and the radio continuum. The radio continuum tail was found to have an extent of $\approx50$\,kpc \citep{Gavazzi1995peculariA1367,Roberts2021LoTSSJellyfishGalaxies}. In our observations, we measure its size to be $ 135\,$kpc, making it the second-longest known next to UGC\,6697. The disk spectral index is $\alpha=-0.71\pm0.08$.  The spectrum quickly steepens to $\alpha \sim-1.1$ outside of the galaxy, and then stays roughly constant until it flattens again to $\approx-0.8$ after $\sim80\,$kpc, in a region that also shows a re-brightening (as shown in \autoref{sec:appendix_spec}). 

\begin{figure}
    \centering
    \includegraphics[width=0.5\linewidth]{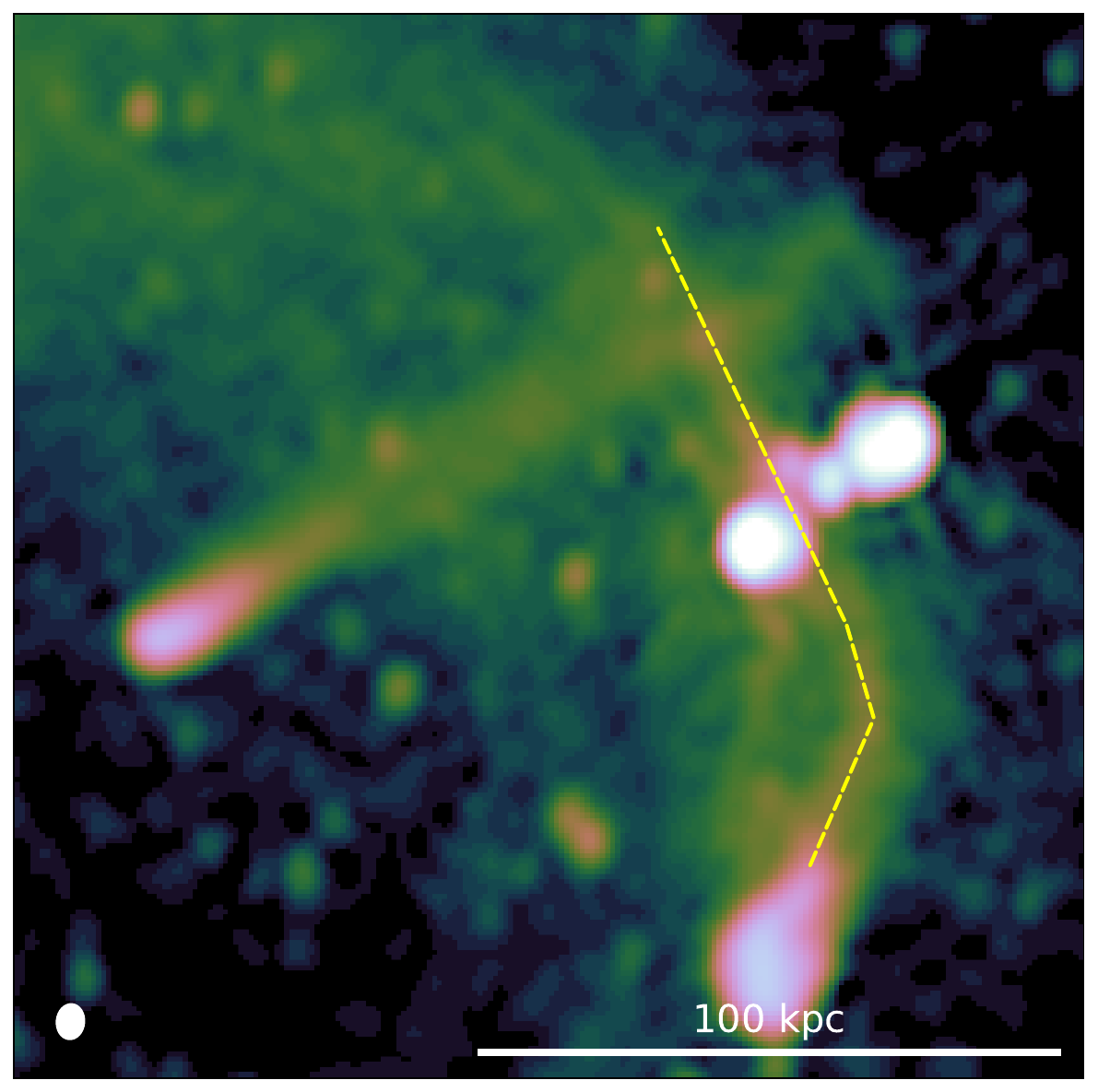}
    \caption{MeerKAT 817\,MHz map of C\,097 and C\,073 at $13''$ resolution. The yellow line guides the eye along the tail of C\,073 and its speculative northern extension. }
    \label{fig:cgcg097}
\end{figure}

\subsection{CGCG 097-073}
This galaxy shows long tails in the ionized gas and the radio continuum, the latter was found to have an extent of at least 75\,kpc in the literature \citep{Gavazzi1995peculariA1367,Boselli2014OriginFaintendRed,Roberts2021LoTSSJellyfishGalaxies,Pedrini2022MUSESneaksPeek}. Due to the lower surface brightness of the tail and the blending with the diffuse emission of the relic and a bright background radio galaxy, the true extent of this tail 
remains unclear from LOFAR observations. Notably, the main component of the tail appears offset to the west from the primary component of the ionized gas tail \citep{Gavazzi200175KiloparsecTrails,Boselli2014OriginFaintendRed,Yagi2017,Pedrini2022MUSESneaksPeek}. 
A filament-like structure in the radio maps, best visible in the MeerKAT UHF band observation displayed in \autoref{fig:cgcg097}, may be interpreted as a continuation of the radio tail toward the north. This is also in agreement with the most distant section of the tail observed in the H$\alpha$ \citep{Scott2015HighlyPerturbedMolecular}.

The disk of C\,073 has a spectral index of $-0.60\pm0.08$, the spectrum of the tail is only slightly steeper, with $\alpha=-0.77\pm0.08$. For this galaxy, there is only a mild spectral steepening visible along the tal, followed by a flattening after 50\,kpc. It is by far the flattest tail of the three, even in the more distant, tentative extension of the tail, $\alpha\geq-1.0$.

\section{Spectral analysis of the tails}\label{sec:tailanalysis}
In SF galaxies, the emission at frequencies of several gigahertz is dominantly due to synchrotron emission. At $\sim1$\,GHz and below, the contribution of thermal emission is $\leq10\%$ \citep{Condon1992}. The radiating CRe are primarily injected at supernova shocks with a power-law energy distribution following an energy index of $\delta\approx -2$. This corresponds to synchrotron radiation with a spectrum of $\alpha\approx-0.5$ \citep{Drury1983}. In the absence of further acceleration processes, the energy distribution of the injected CRe is modified only by energy loss  processes: in the presence of magnetic and photon fields, synchrotron and inverse-Compton losses will more efficiently deplete high-energy CRes emitting at high frequencies leading to a steepening of the spectrum known as spectral aging \citep[e.g.,][]{Harwood2013}. In the dense ISM, energy losses via Bremsstrahlung and ionization are also relevant  \citep[see e.g.,][]{Basu2015}. Within the disks, the balance of these processes typically leads to spectral indices $\sim -0.6$, with more massive galaxies showing steeper spectra \citep[e.g.,][]{Heesen2022NearbyGalaxiesLOFAR,Edler2024ViCTORIAProjectLOFAR}.

Spectral aging models are commonly employed to determine the radiative age of the synchrotron-emitting plasma of radio galaxies, where loss processes with the thermal gas or galaxy radiation field can be ignored for the approximately giga-electronvolt electrons that we probe in the radio band \citep[e.g.,][]{Kardashev1962NonstationaritySpectraYoung,longair2010high,Harwood2013,Harwood2015SpectralGalaxies,Edler2022Abell1033Radio}. This should also hold for the low-density tails of ram pressure-stripped galaxies in cases where there is no star formation, as argued by \citet{Ignesti2023RadioContinuumTails}. 
This justifies recent approaches to model the synchrotron tails of RPS galaxies in an analogous way \citep{Ignesti2023RadioContinuumTails,Roberts2024RadiocontinuumSpectraRampressurestripped,Roberts2024NGC2276}. Typically, it is found that the tails are well modeled by the classical spectral aging model of \citet{Jaffe1973} (hereafter JP-model). 
This model assumes that CRe in a certain region were all injected in a single instantaneous burst with some injection electron index $\delta_\mathrm{inj}$ and subsequently, experienced synchrotron and inverse Compton losses in a uniform magnetic and radiation field. The JP model assumes that the CRe isotropize rapidly due to frequent scattering, such that all CRe effectively synchrotron-radiate with an isotropic pitch angle distribution. Given the lifetime of the observed CRe of $t>10^7$\,yr, frequent scattering is expected, making the model more realistic compared to the models of \citet{Kardashev1962NonstationaritySpectraYoung,Pacholczyk1970} which assume constant pitch angles across the CR lifetime. 
In the JP-model, energy losses introduce a spectral break in the CRe spectrum and consequently, in the observed synchrotron emission. The connection between the break frequency $\nu_\mathrm{b}$, the age $t$, and the  magnetic field strength $B$ is given by the following equation \citep[e.g.,][]{Murgia1999SynchrotronSpectraAges}:
\begin{equation}\label{eq:lifetime}
    t_\mathrm{rad} =  3.2\times 10^4 \frac{\sqrt{B}}{B^2+B_\mathrm{CMB}^2}\times \frac{1}{\sqrt{\nu_b}}\,\mathrm{Myr},
\end{equation}
where $B$ is measured in $\mu$G and $\nu_b$ in megahertz. Here, inverse Compton scattering is assumed to be only with the cosmic microwave background (CMB), and $B_\mathrm{CMB}\approx3.25(1+z)^2\,\upmu$G is the magnetic field strength equivalent to the CMB energy density.

For the spectral-aging model of the tails to hold, the injection of fresh CRe into the tail due to star formation needs to be negligible compared to the stripped CRe. While C\,073 and C\,079 show no signs of star formation in the tails \citep{Pedrini2022MUSESneaksPeek}, UGC\,6697 shows trails of \hii{} regions in the initial $\approx40$\,kpc of the tail, extending beyond the stellar disk, as we show in \autoref{fig:si_tails} \citep{Consolandi2017UGC6697}.
There are no clear morphological or spectral features in the radio maps corresponding to the location of the \hii{} regions. However, we cannot exclude the possibility that for this first, shorter section of the tail right behind the stellar disk, star formation partly contributes to the observed radio emission. Nevertheless, there are no signs of star formation taking place in the final 280\,kpc of the tail.

\subsection{The pure-aging scenario}\label{sec:agingonly}
If the spectrum of the tails is shaped only by radiative losses, the spectral index is expected to steepen monotonically along the tail away from the disk. For UGC\,6697, this is the case only for the initial part of the tail; afterward, the spectral index displayed in \autoref{fig:si_tails}  stays constant and even increases again (corresponding to a flattening of the spectrum) beyond $r=200\,$kpc. 
Also for the other two galaxies, this monotonic steepening is not observed, as can be seen in \autoref{sec:appendix_relicmodel}.

Spectral aging should also lead to a characteristically curved spectrum due to the quicker depletion of high-energy electrons.  
In \autoref{sec:appendix_spec}, we show the color-color diagrams $\alpha_{ij}$ versus $\alpha_{kl}$ for the three galaxies. They were created based on the four frequencies considered in this work and using the regions of \autoref{fig:tailregs}. 
We also compare them to the JP aging model, which we generated using \texttt{synchrofit}\footnote{\url{https://github.com/synchrofit/synchrofit}} \citep{Quici2022SelectingModellingRemnant}.
In these plots, the initial data points close to the galactic disks, at distances of a few 10\,kpc, are in agreement with the JP-model (within uncertainties). However, further along the tail, where the spectrum steepens, the data points disagree with the model. In particular, for UGC\,6697, where we have the most data points, the spectrum does not show the expected curvature, but a more even steepening across all observed frequencies. 

Lastly, the extent of the longest tail, UGC\,6697, is challenging to explain in a scenario with only energy losses, as we argue in the following: the radiative lifetime, $t_\mathrm{rad}$, in \autoref{eq:lifetime} takes its maximum value for the minimum loss magnetic field $B=B_\mathrm{min}=B_\mathrm{CMB}(z)/\sqrt{3} \approx 1.93\,\mathrm{\upmu{G}}$. Therefore, assuming $B=B_\mathrm{min}$ allows us to place an upper limit on $t_\mathrm{rad}$.
The frequency of the spectral break, $\nu_\mathrm{b}$, reaches the central frequency of our highest observed band of 1.27\,GHz after $t_\mathrm{rad}\approx80\,$Myr.
Thus, for the most distant part of the tail at $r=300$\,kpc to be stripped 80\,Myr ago, the galaxy would need to travel at a projected velocity of $\geq3660\,\mathrm{km\,s^{-1}}$. 

We compare this to a dynamical estimate of the maximum possible projected velocity for UGC\,6697: assuming the galaxy is gravitationally bound to the cluster, its velocity in three dimensions cannot exceed the escape velocity: $v_\mathrm{3D} \leq v_\mathrm{esc}(r)$. 
The escape velocity at a distance, $r$, from the cluster center can be calculated according to
\begin{equation}
    v_\mathrm{esc}(r) = \sqrt{\frac{2GM(r)}{r}}
,\end{equation}
in the case of a spherical cluster mass distribution, $M(r)$. Here, $G$ is the gravitational constant. Approximating the mass distribution using the profile of \citet{Navarro1996StructureColdDark} as fit for A1367 by \citet{Rines2003CAIRNSClusterInfall}, we get \citep{Lokas2001PropertiesSphericalGalaxies}
\begin{equation}
    M(r) = M_{200}\frac{f\left(\frac{cr}{r_{200}}\right)}{f(c)}; \qquad f(x) = \ln\left(1+x\right) - \frac{x}{1+x}.
\end{equation}
with the concentration parameter $c=16.9$.
The distance between the hierarchical cluster center determined by \citet{Rines2003CAIRNSClusterInfall} and UGC\,6697 is at least the projected distance of 446\,kpc, yielding $v_\mathrm{esc} \leq 2075\,\mathrm{km\,s^{-1}}$. Given the $491\,\mathrm{km\,s^{-1}}$ difference in line-of-sight velocity compared to the northwestern brightest cluster galaxy (BCG) NGC\,3842 \footnote{Line-of-sight velocities were taken from the NASA/IPAC Extragalactic Database \url{https://ned.ipac.caltech.edu/}.}, the infall velocity of UGC\,6697 in the plane of the sky should not exceed 2017\,km\,s$^{-1}$, in conflict with the minimum velocity estimate of $\geq3660\,\mathrm{km\,s^{-1}}$ obtained from the size of the observed tail.
If there are substantial bulk motions within the ICM and it is not at rest with respect to the cluster center, this discrepancy may decrease; see the discussion in \autoref{sec:discussion}.

In the following, we consider several scenarios that can explain the observed non-monotonous spectral trends and the extreme linear size of the UGC\,6697 tail. The aim of this is to investigate which physical processes can boost the observed emission of the distant part of the tail while also flattening the observed synchrotron spectrum and reducing the observed spectral curvature.

\subsection{The shock-compression scenario}\label{sec:comp}
\begin{figure}
    \centering
    \includegraphics[width=0.99\linewidth]{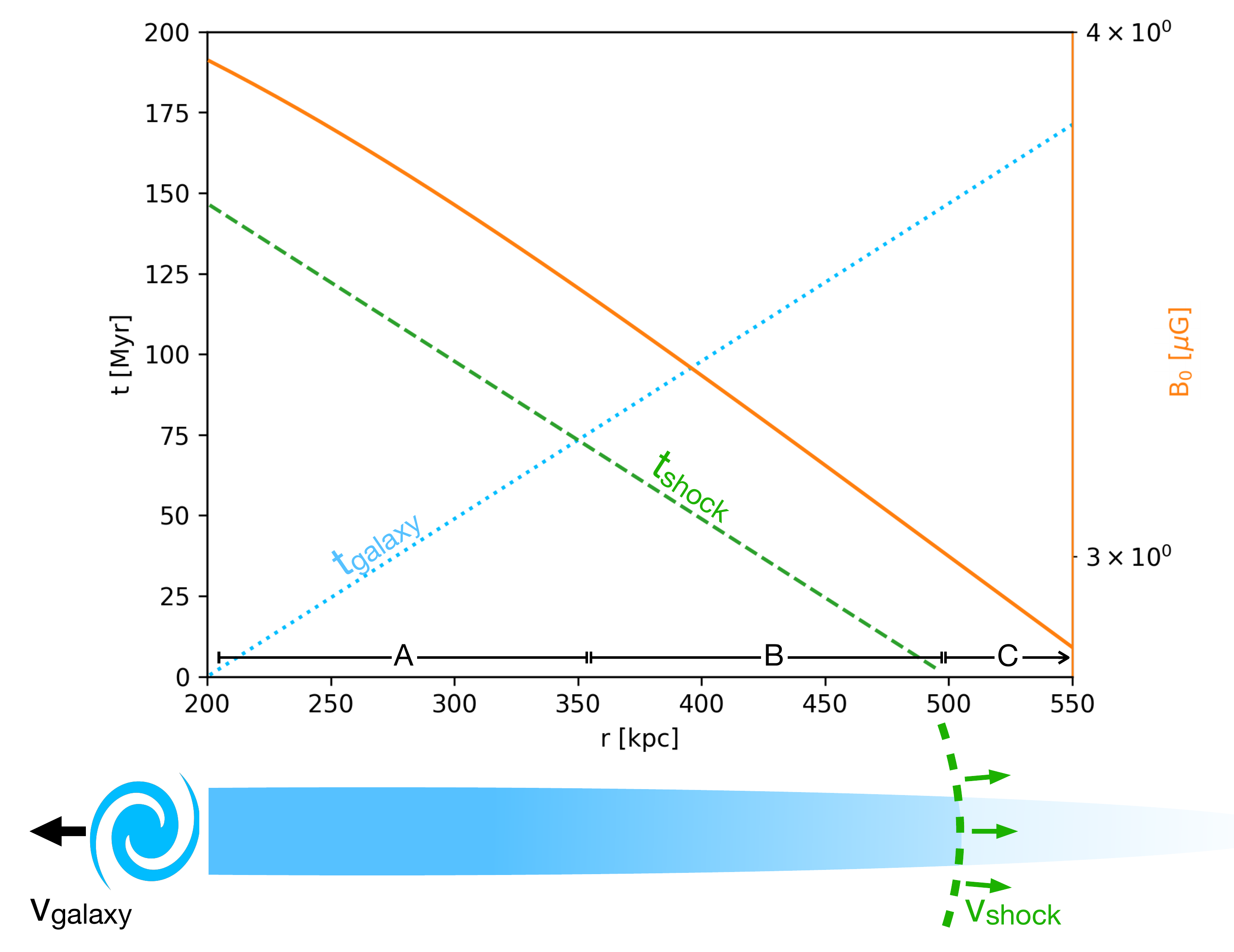}
    \caption{Time since the galaxy (blue) and shock (green) passed at a radius of $r$ (left $y$ axis). The right $y$ axis shows the magnetic field (orange) as function of radius. The sketch below highlights the present situation, with the galaxy propagating to the left and the shock front to the right. Regions A, B, and C correspond to those defined in \autoref{sec:comp}}
    \label{fig:sketch}
\end{figure}
The first scenario that we consider that could boost the observed emission while also leading to a spectral flattening is the amplification of the tail magnetic field by compression due to the large-scale ICM shock reported in \citet{Ge2019MergerShockAbell}. This scenario is also motivated by the fact that the end of the UGC\,6697 tail is in close proximity to the ICM shock (the same holds for C\,079). To model this scenario, we assumed that the galaxy falls radially toward the cluster center with a velocity on the order of its escape velocity $v_\mathrm{galaxy}=-2000\,\mathrm{km\,s^{-1}}$, leaving behind a tail that is quickly decelerated to be at rest with respect to the ICM and cluster center. The shock front moves in the opposite direction with $v_\mathrm{shock}=2000\,\mathrm{km\,s^{-1}}$, approximately the velocity corresponding to the radio Mach number $\mathcal{M}_\mathrm{r}=\sqrt{\frac{\alpha+1}{\alpha-1}}\approx2.5$ and the speed of sound of $\mathrm{c_\mathrm{s}}=840\,\mathrm{km\,s^{-1}}$ \citep{Ge2019MergerShockAbell}.
Here, we assumed the radio Mach number, since using the smaller Mach number $\mathcal{M}_x=1.6$ measured from the X-rays, the compression will only have a weak effect.
The discrepancy between $\mathcal{M}_x$ and $\mathcal{M}_r$ can be explained with the X-ray measurement being affected by projection effects and by the two wavelengths tracing different parts of the Mach-number distribution \citep{Botteon2020ShockAccelerationEfficiency,Wittor2021ExploringSpectralProperties}.
For simplicity, we assumed the shock and galaxy velocities to be constant. The ICM was modeled to follow a $\beta$-model \citep{Cavaliere1976XraysHotPlasma} with $\beta=2/3$ and a critical radius of $r_\mathrm{c}=500$\,kpc. The magnetic field at the critical radius was assumed to be $B_c=3\,\mathrm{\mu{G}}$ and to scale with the ICM density as $B_0\propto n^{1/2}$  \citep{Bonafede2010ComaClusterMagnetic} .
At present time, the start of the tail is at $r_\mathrm{galaxy}=200$\,kpc and the shock front is at $r_\mathrm{shock}=480\,$kpc. 
At a radius $r$, the galaxy passed a time of $t_\mathrm{galaxy}=(r-r_\mathrm{galaxy})/v_\mathrm{galaxy}$ ago, while the shock passed at $t_\mathrm{shock}=(r-r_\mathrm{shock})/v_\mathrm{shock}$. This situation, together with the magnetic field model, is displayed in \autoref{fig:sketch}.

The impact of the shock on the tail in our model is twofold. Firstly, the passage of the shock with Mach-number $\mathcal{M}$ compresses the magnetic field, $B$, in the tail. For the compression of an unordered magnetic field, $B\propto n^{2/3}$. Thus, the magnetic field compression factor, $f_B$, can be derived from the density jump, $C$, of the Rankine-Hugoniot jump conditions for a monoatomic gas \citep{Landau1959FluidMechanics}:  
\begin{equation}\label{eq:comp}
    C = \frac{4\mathcal{M}^2}{\mathcal{M}^2+3}
,\end{equation}
according to $f_B = C^{2/3}$.
Secondly, we assume that the adiabatic invariant $p^2_\perp/(Bm_e)$ is conserved across the shock, where $m_e$ is the electron mass and $p_\perp$ the momentum. This would cause the CRe to be heated according to the model of \citet{Ensslin2001RevivingFossilRadio} (see also \citealt{Markevitch2005BowShockRadio}), effectively shifting the spectrum to higher energies by $E\propto B^{1/2}$. If this is indeed the case will depend on the exact properties of the shock and ICM. Both of these effects can increase the synchrotron brightness and flatten the observed spectrum, since we would observe the CRe spectrum at lower energies, where it is less curved. 
This model leads to three distinct regions along the tail:
\begin{enumerate}[A:]
\item In the regime where $t_\mathrm{galaxy}{<}t_\mathrm{shock}$, the CRe spent their full life in the compressed magnetic field and were never heated by the passing shock. 
\item In the region $t_\mathrm{galaxy} > t_\mathrm{shock}$ \& $t_\mathrm{shock} >= 0$, the electrons spent part of their lifetime in the uncompressed magnetic field, thereby aging less quickly, and were then spun up in energy by the shock. 
\item Lastly, the region that the shock did not yet pass ($t_\mathrm{shock} < 0$), such that the CRe radiate only in the uncompressed magnetic field and were never spun up in energy.
\end{enumerate}
These regions are also marked in \autoref{fig:sketch}.
We model the tail CRe spectrum $N(E)$ by evolving the following simplified Fokker-Planck equation numerically for the time, $t_\mathrm{galaxy}$, at each radius, $r$: 
\begin{equation}\label{eq:fokkerplanck}
    \frac{\partial N(E,t,r)}{\partial t} = \frac{\partial}{\partial E} \left[ N(E,t,r)b(E,r,t)\right].
\end{equation} 
Here, the loss term $b(E,t,r)$ captures losses due to synchrotron and inverse Compton with the CMB is given by 
\begin{equation}
b=6.4\times 10^{-6} \mu_0 \left(B(r,t)^2 + B_\mathrm{CMB}^2\right) E^2\,\mathrm{m^3\,MeV^{-2}\,Myr^{-1}}.
\end{equation}
The time dependency in $B(r,t)$ is due to the shock compression of the magnetic field at a time of $t_\mathrm{shock}$. 
 Our model neglects CR diffusion and losses due to the adiabatic expansion of the tail.
The adiabatic heating of the CRe is implemented  by implicitly shifting the energy spectrum to higher energies by $E \propto B^{1/2}$ at the time, $t_\mathrm{shock}$, where the shock passes at $r$. 
We assume the initial electron distribution to follow a power law with an energy index of $\delta = -2.6$, corresponding to the observed synchrotron slope of $\alpha \approx -0.8 $ observed at the trailing edge of the galaxy.
The synchrotron emission of $N(E)$ in the magnetic field, $B(r)$, was calculated using standard synchrotron equations. More details of the model are reported in \autoref{sec:appendix_CRmodel}.

\begin{figure}
    \centering
    \includegraphics[width=0.95\linewidth]{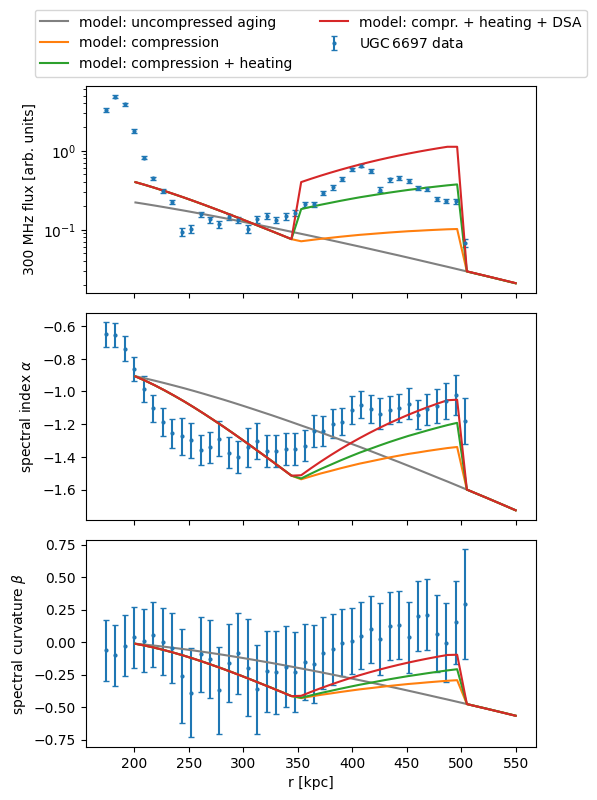}
    \caption{Comparison between the observations and our model assuming aging without compression or acceleration (gray), only magnetic field compression (orange), magnetic field compression and adiabatic heating (green), and magnetic field compression and shock reacceleration (red).
    The top panel shows the flux density fit at 300\,MHz, the center panel the spectral index, and the bottom panel the spectral curvature.}
    \label{fig:model_comp}
\end{figure}

We evaluate the synchrotron emission of the models at the four observed frequencies and then fit \autoref{eq:logparabola} to those values around the central frequency $\nu_0=300\,$MHz to obtain the flux density, spectral index, and spectral curvature. 
We compare the observed data of UGC\,6697 to the model without a shock and without compression, the model with only magnetic field compression and to the model with magnetic field compression and adiabatic CR heating in \autoref{fig:model_comp}. We did not attempt to model the contribution of the ISM magnetic field, which is required to accurately describe the emission of the disk and the initial tens of kiloparsecs of the tail.

Clearly, shock compression of the magnetic field can increase the brightness, flatten the spectral index and bring the curvature closer to the power-law case $\beta=0$ for the distant part of the tail. If the CRe are also adiabatically heated by the shock, this effect is amplified. However, even with the adiabatic heating and assuming the higher radio Mach number instead of the lower X-ray one, the model cannot explain the observed spectrum in region B, which is both rather flat $\alpha\sim -1.0$ and straight (in log-space, i.e., a power law) or even convex. We could not find any reasonable combination of model parameters that closely matches the observed flattening.

\subsection{The diffusive shock acceleration scenario}\label{sec:re-accel}
In addition to the compression, the shock front could also reaccelerate the CRe via DSA or turbulent (Fermi-II) reacceleration processes, with the stripped CRe in the tail being the seed population. Such scenarios are also employed to explain the peculiar spectral properties of a number of tailed radio galaxies in clusters \citep[e.g.,][]{Lusetti2024ReenergizationAGNHeadtail}. Here, we only consider DSA as an acceleration mechanism, given its stronger theoretical foundation and the presence of the radio relic, which is a  signature of shock-acceleration in the ICM.

Diffusive shock acceleration enters \autoref{eq:fokkerplanck} as acceleration term $a(E)={\mathrm{d}{E}}/{\mathrm{d}t}=E/t_\mathrm{acc}(r,t)$, leading to the following modified equation:
\begin{equation}
    \frac{\partial N(E,t,r)}{\partial t} = \frac{\partial}{\partial E} \left[ N(E,t,r) \,\{b(E,r)-a(E,r,t)\}\right].
\end{equation} 
Effectively, for each radial cell, we use a finite $t_\mathrm{acc}$ only during the time interval in which the shock propagates through that cell; otherwise, the term vanishes. We assumed a value of $t_\mathrm{acc}=5$\,Myr, compared to a time of 4.3\,Myr the shock takes to cross a radial cell. The curve corresponding to the DSA scenario is shown in \autoref{fig:model_comp}. More efficient acceleration leads to a drastic increase in surface brightness for the shocked part of the tail, which is not seen in the observed data to that extent.

Overall, the model with standard DSA reproduces the main features of the observed radio emission and is well motivated by the presence of the X-ray shock and the radio relic. Our concrete choice of model falls only slightly short in reproducing the observed curvature. Independent of the parameters, it cannot reproduce the slightly convex spectrum observed in some regions of the final part of the tail, with values of $\beta > 0$.

We note that  our models do not aim to reproduce the rapid initial decline in surface brightness observed in the transition from the disk to the tail. This is likely due to the stronger magnetic field within the disk ISM, which is not included in our simplistic model.
Nonetheless, our analyses indicates that the observed surface brightness increase farther along the tail, and subsequent flattening of the spectrum, can be explained by the passage of a shock.

\section{The radio filaments surrounding NGC 3842}\label{sec:filament}
\begin{figure}
    \centering
    \includegraphics[width=1.0\linewidth]{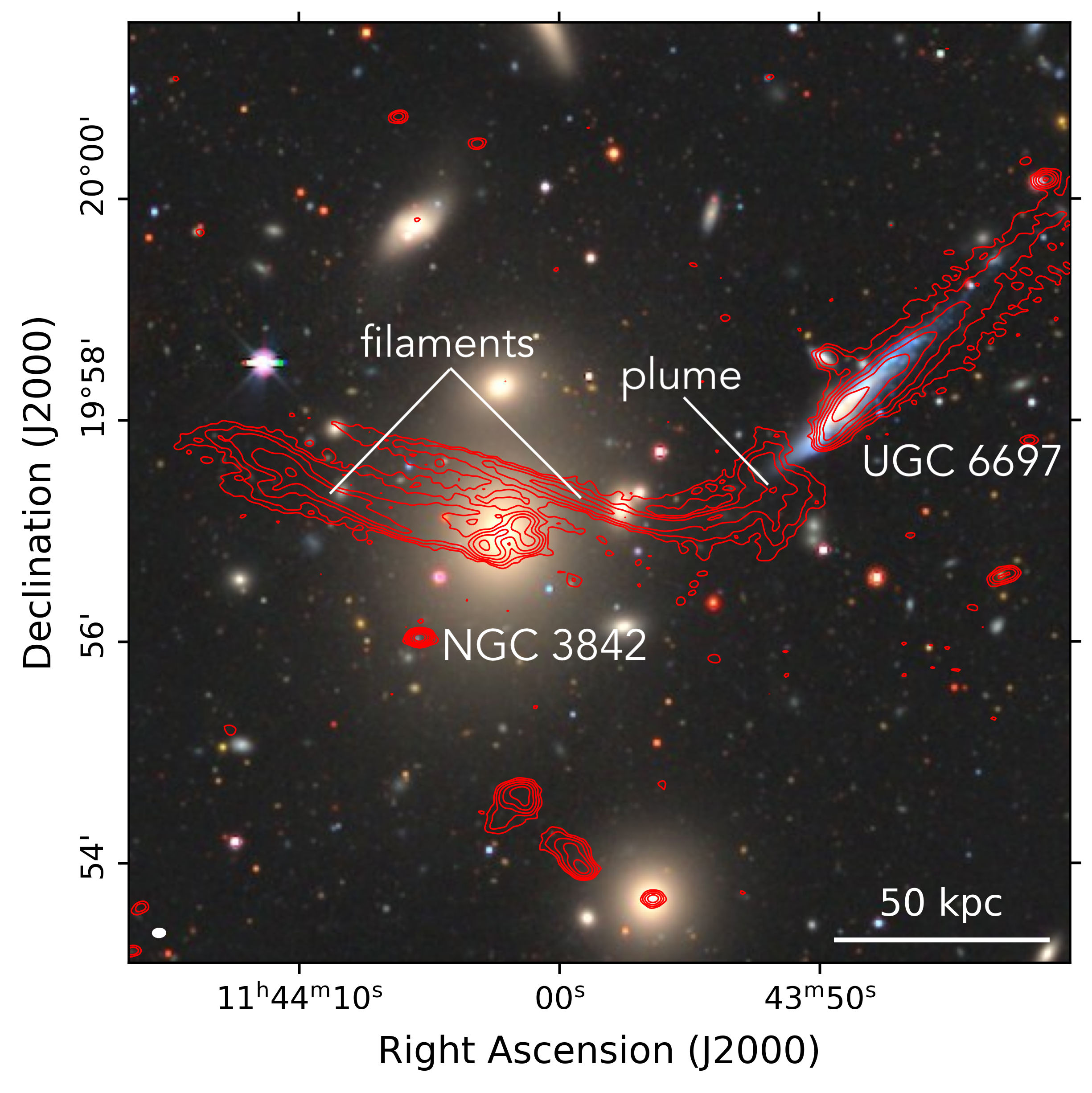}
    \caption{Optical image of NGC\,3842 and surroundings from the DESI Legacy Survey DR 10 \citep{Dey2019}. Overlaid in red are the LOFAR HBA contours, taken from an image with a PSF of $6''\times4''$. The contours are $[4,8,16,32,64]\times \sigma_\mathrm{rms}$, where $\sigma_\mathrm{rms}= 77\,\mathrm{\mu{Jy}\,beam^{-1}}$.}
    \label{fig:paralleltails}
\end{figure}

NGC\,3842, the BCG of the northwestern subcluster, shows a double-lobed morphology of 20\,kpc extent with a typical radio galaxy spectral index of $\alpha\approx-0.7$. Beyond this central emission, the galaxy is surrounded by two parallel tails, as shown in \autoref{fig:paralleltails}. Only the southern source is connected to NGC\,3842, initially starting as a bifurcated filament starting from the two lobes of NGC\,3842, which then transforms into a single one 30\,kpc to the east. The largest linear size of the emission is 140\,kpc, the narrowest part of the filaments appears to be below 4\,kpc in size. The filament appears to show even narrower filamentary sub-structures.

The radio spectrum of the filaments is highly peculiar, being both ultra-steep and strongly curved (see \autoref{fig:fluxes}). The spectrum steepens from $\alpha_{54}^{144}=-1.57\pm0.04$ at LOFAR frequencies to $\alpha_{817}^{1270}=-2.48\pm0.2$ at MeerKAT frequencies.
The spectrum is also remarkably uniform across the source (see \autoref{fig:si}).
Both the morphology of the filaments and the spectrum do not resemble typical radio galaxies. It does, however, share properties with a number of filamentary radio sources, typically connected to the radio lobes of BCGs \citep{Ramatsoku2020CollimatedSynchrotronThreads,Rudnick2022IntraclusterMagneticFilaments,Macgregor2024EvolutionaryMapUniverse}, discovered in recent years. 
A possible explanation for this was recently raised by \citet{Churazov2025}, suggesting that such structures are magnetic filaments, along which CRe are able to propagate significantly faster along pressure gradients compared to if they were bound to the thermal gas. 
To add to the puzzle, the northern filament ends in a plume directly, at least in projection, in front of the stellar disk of UGC\,6697. The plume is separated from the contact discontinuity on the leading edge by approximately 5\,kpc, as measured from the high-resolution LOFAR HBA map. The plume itself could be interpreted as compression of the filament by the infalling UGC\,6697. If the filaments both originate from NGC\,3842, it appears unlikely for one of them to end exactly toward the infalling direction of UGC\,6697. 

If the mechanism proposed by \citet{Churazov2025} is indeed responsible for the observed filaments, a possible scenario, also able to explain the apparent connection to UGC\,6697, would resemble the following: the central part of the NGC\,3842 subcluster contains multiple magnetic filaments that are oriented approximately in the east-west direction, possibly due to former AGN activity. Along the southern filament, CRe initially accelerated by the central AGN can move rapidly. They mix with CRe populations of various radiative ages, yielding the observed spectrum. Similarly, the northern filament could be explained by CRe originating from the plume at the leading edge of UGC\,6697 that are radio-brightened at the Mach cone of the galaxy by DSA and/or compression and then transported east.
While this scenario is highly speculative, it could explain the enigmatic connection between the northern filament and UGC\,6697.

\section{Discussion}\label{sec:discussion}
The three galaxies that are subject of this work and their tails are examples of extreme RPS in multiple ways: they show the longest known radio continuum tails,  they have a strong radio excess given their SFR, and their tails show peculiar, non-monotonic spectral trends.  
This raises the question of why three objects in close proximity to each other show these peculiarities that distinguish them from the general population of RPS galaxies. 
The likely explanation for this lies in the ICM shock that (at least in projection) must have crossed all three galaxies. The ICM shock can explain the extreme RPS by introducing a density jump and  ICM bulk motion. It is further supported by the similar orientation of the tails, which approximately follow the propagation direction of the shock.
The increase in density given by \autoref{eq:comp} would be in the range of $\mathcal{C}\approx1.8 \mathrm{-} 2.7$ when considering the X-ray and radio Mach number, respectively. 
Approximating the ICM as an ideal gas, the velocity of the bulk motion following the shock is given by
\begin{equation}
\Delta{v} = \frac{3}{4}c_\mathrm{s}\mathcal{M}\left(1-  \frac{1}{\mathcal{M}^2}\right).  
\end{equation}
The resulting velocity of the bulk motion will be $614$-$1323\,\mathrm{km\,s^{-1}}$ based on the Mach number assumption. 
Depending on the geometry and the infall velocity, the ICM bulk motion may temporarily increase the relative velocity between the infalling galaxy and the ICM by a factor of $1+\Delta{v}/v_\mathrm{infall}$. 
For UGC\,6697 and C\,079, the propagation direction of the shock is antiparallel to the projected velocity of the galaxies, making this scenario likely. The resulting increase in ram pressure, $P\propto \rho{v}^2$, would be a factor of $\mathcal{C}(1+\Delta{v}/v_\mathrm{infall})^2$. For $\mathcal{M}=2.5$ and $v_\mathrm{infall}=2000\,\mathrm{km\,s^{-1}}$, the temporary increase in ram pressure would be a factor of up to 7.
Thus, the encounter with an ICM shock is able to substantially enhance RPS. At the same time, it can also explain the peculiarities of the radio tails, as we argued in \autoref{sec:re-accel}.

Clusters with radio relics, tracers of ICM shocks, are moderately frequent (e.g., $\sim10\%$ occurrence rate in the sample of \citealt{Botteon2022PlanckClustersLOFAR}) with many tens of examples known from radio and X-ray studies. If the extreme stripping events and radio tails in A1367 are caused by the ICM shock, that raises the question of why such interactions are not observed in other clusters with shocks.
Due to the limited angular resolution of current large-area surveys, known radio tails behind SF galaxies are mostly limited to low-redshift systems of $z\lesssim0.05$. Conversely, only very few systems with ICM shocks fall into this range - none out of the 309 clusters in the sample of \citet{Botteon2022PlanckClustersLOFAR}. Among the few examples other than A1367 are Coma (A\,1656), A\,3376, and A\,3667. For Coma and A\,3376, no direct interactions between SF galaxies and shocks are known. However, Coma shows an unusually large number of galaxies with prominent stripped tails \citep{Roberts2024RadiocontinuumSpectraRampressurestripped}, and A\,3376 hosts a population of post-starburst galaxies, possibly originating from past interactions with the shock \citep{Kelkar2020PassiveSpiralsShock}.
For A\,3667, a visual inspection of deep MeerKAT images (\citealt{deGasperin2022MeerKATViewDiffuse}, Benati+ in prep) indeed revealed a connection between the jellyfish galaxy WISEA J201014.69-563830.1 and the western radio relic. This galaxy has an outer ring that mostly contains stars formed in the last 500\,Myr. An encounter with the ICM shock may reasonably have been the trigger. Such a scenario was already suggested by \citet{Moretti2018GASPRampressureStripping}.
Thus, interactions between ICM shocks and SF galaxies may be understood as a more important trigger of RPS and as a source of seed electrons giving rise to diffuse radio emission in merging clusters once higher resolution radio surveys become available with telescopes such as LOFAR\,2.0 and the SKA. 

Next, we want to highlight certain limitations of our model for the tail of UGC\,6697 and discuss its applicability to C\,073 and C\,079.
One factor of uncertainty is the magnetic field structure in the disk and tail. We did not attempt to model the contribution of the ISM magnetic field, which is necessary to explain the initial 50\,kpc of the emission, particularly within the disk.  Also, if the tail magnetic field is strongly aligned with the tail, as is suggested by the magnetic draping scenario \citep[and references therein]{Muller2021NatAst}, the compression factor of the tail magnetic field may be less compared to the unordered field that we assumed. Furthermore, our model assumes that all radio emission at a certain distance from the galaxy originates from CRe that were injected at a single time. In practice, since CRe are injected in an extended region in the disk and the initial 30\,kpc of the tail, aged CRe that were injected at earlier times close to the leading edge of the galaxy will be advected along the disk and tail where they will mix with more recently injected CRe. This mixing may reduce the observable spectral curvature.
Lastly, when connecting the infall velocity of the galaxy with respect to the cluster center to the relative velocity between the galaxy and the ICM, we assumed that ICM bulk motions are negligible. Especially in a merging cluster and behind the shock front, such bulk motions may be considerable, as we argued earlier in this section. Thus, the relative velocity between the ICM and the galaxy may be greater than its infall velocity onto the cluster. This may weaken our point on the maximum allowed relative velocity of the galaxy of  \autoref{sec:agingonly}.

Lastly, we briefly want to discuss how our model for UGC\,6697 applies to the other two RPS galaxies in the relic region. For C\,079, given its similar (projected) direction to UGC\,6697, the model could be applied in an analogous way to explain the lack of spectral steepening and in particular the spectral flattening in the final section of the tail.
For C\,073, the encounter between the shock and the tail may have been closer to perpendicular, meaning that reacceleration along the tail might have taken place more recently across a shorter time frame. This is in agreement with the observed flat spectrum of the tail.

\section{Conclusions}\label{sec:conclusion}
In this work, we presented novel LOFAR and MeerKAT observations of the nearby merging cluster A1367, focusing on three RPS galaxies (UGC\,6697, C\,073, and C\,079) located near an ICM shock and radio relic. The relic itself is the subject of an accompanying work (Hoeft et al. in prep.). A secondary subject of our study is a newly detected, enigmatic filamentary source at the leading edge of UGC\,6697.

Our findings are:
\begin{itemize}
    \item The radio tails are significantly longer than previously reported.  UGC\,6697 shows a $300$\,kpc tail, the longest known for a SF galaxy at any wavelength.
    \item The tails are morphologically distinct from the diffuse radio relic emission, as we confirmed by subtracting a model of the relic emission. 
    \item The integrated disk spectral indices are in the typical range for SF galaxies ($\alpha= -0.6$ to $-0.7$), while the tails are significantly steeper ($\alpha= -0.8$ to $-1.2$). 
    \item We confirm that the radio luminosity of the galaxies is far in excess of the extrapolation from their SFR based on scaling relations for normal SF galaxies. 
    \item For UGC\,6697, the tail exhibits an unexpected re-brightening and spectral flattening after $150\,$kpc. 
    Further, the extreme tail length requires galaxy velocities significantly larger than the escape velocity estimates. Thus, we rule out a pure-aging scenario. Similar non-monotonous spectral variations are observed for C\,073 and C\,079.
    \item We present a model for UGC\,6697  that includes an encounter between the stripped tail and the ICM shock. Via magnetic field compression  and diffusive shock reacceleration of the stripped cosmic ray electrons, this can explain the length, the re-brightening, and the spectral flattening of the tail. 
    \item We discover two narrow ($\leq 4\,$kpc), ultra-steep ($\alpha=-1.74\pm0.07$), and extremely curved ($\beta=-0.6\pm0.26$) filamentary structures.  One of them terminates in a plume at UGC\,6697's leading edge; this may represent the first tentative detection of particle energetization at a galaxy's bow shock. The filaments may trace magnetically dominated structures in the cluster center.
\end{itemize}
These results represent direct evidence that ICM shocks in merging clusters affect the properties of their member galaxies, while simultaneously reenergizing stripped cosmic rays. 
Future high-resolution radio surveys with LOFAR 2.0, MeerKAT+, and SKA will reveal how common such shock-galaxy interactions are and  what role they play in accelerating galaxy evolution in cluster environments.
\begin{acknowledgements}
FdG acknowledges support from the ERC Consolidator Grant ULU 101086378. M.B. acknowledges support from the Deutsche Forschungsgemeinschaft under Germany's Excellence Strategy - EXC 2121 “Quantum Universe” - 390833306 and from the BMBF ErUM-Pro grant 05A2023. AI acknowledges funding from the European Research Council (ERC) under the European Union's Horizon 2020 research and innovation programme (grant agreement No. 833824). A.D. acknowledges support by the German Federal Ministry of Research, Technology and Space (BMFTR) Verbundforschung under the grant 05A23STA. 
The MeerKAT telescope is operated by the South African Radio Astronomy Observatory, which is a facility of the National Research Foundation, an agency of the Department of Science and Innovation.
LOFAR (van Haarlem et al. 2013) is the Low Frequency Array designed and constructed by
ASTRON. It has observing, data processing, and data storage facilities in several countries, which are owned by various parties (each with their own funding sources), and that are collectively operated by the ILT foundation under a joint scientific policy. The ILT resources have benefited from the following recent major funding sources: CNRS-INSU, Observatoire de Paris and Université d'Orléans, France; BMBF, MIWF-NRW, MPG, Germany; Science Foundation Ireland (SFI), Department of Business, Enterprise and Innovation (DBEI), Ireland; NWO, The Netherlands; The Science and Technology Facilities Council, UK; Ministry of Science and Higher Education, Poland; The Istituto Nazionale di Astrofisica (INAF), Italy. This research made use of the Dutch national e-infrastructure with support of the SURF Cooperative (e-infra 180169) and the LOFAR e-infra group. The Jülich LOFAR Long Term Archive and the German LOFAR network are both coordinated and operated by the Jülich Supercomputing Centre (JSC), and computing resources on the supercomputer JUWELS at JSC were provided by the Gauss Centre for Supercomputing e.V. (grant CHTB00) through the John von Neumann Institute for Computing (NIC).
This research made use of the University of Hertfordshire high-performance computing facility and the LOFAR-UK computing facility located at the University of Hertfordshire and supported by STFC [ST/P000096/1], and of the Italian LOFAR IT computing infrastructure supported and operated by INAF, and by the Physics Department of Turin university (under an agreement with Consorzio Interuniversitario per la Fisica Spaziale) at the C3S Supercomputing Centre, Italy. This work made use of EveryStamp\footnote{\url{https://tikk3r.github.io/EveryStamp/}}. This research has made use of the NASA/IPAC Extragalactic Database, which is funded by the National Aeronautics and Space Administration and operated by the California Institute of Technology.
\end{acknowledgements}
\bibliographystyle{aa}
\bibliography{manual_zotero}
\begin{appendix}
\onecolumn
\section{Radio relic emission model}\label{sec:appendix_relicmodel}
\FloatBarrier
\begin{figure}[h!]
    \centering
    \includegraphics[width=0.28\linewidth]{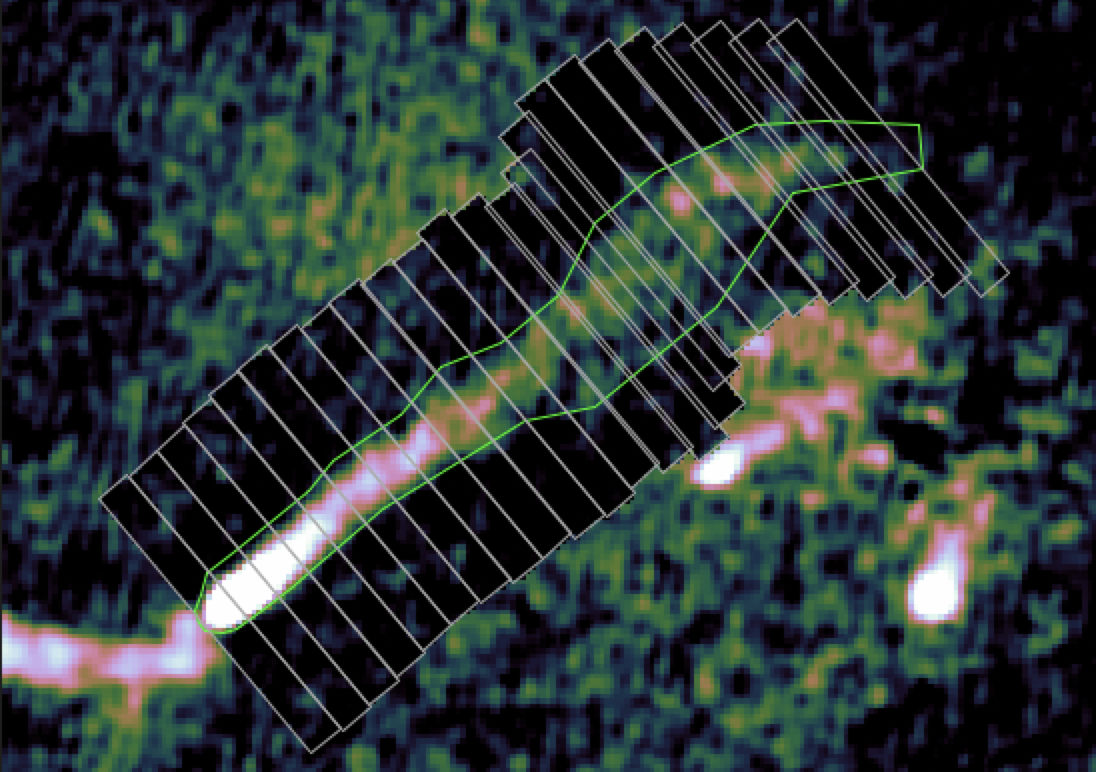}
    \includegraphics[width=0.265\linewidth]{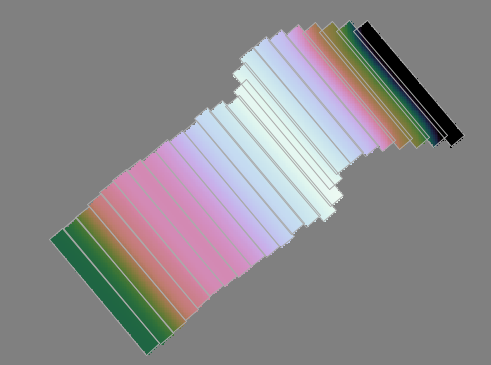}
    \caption{Left: UGC\,6697 at 54\,MHz after subtracting the contribution of the radio relic emission from the tail. The areas inside the gray rectangular boxes but outside the green tail region were used to estimate and subtract the emission of the relic region.
    Right: Subtracted model, with surface brightness ranging from 5\,mJy/beam (white) to 1.2\,mJy/beam (black).}
    \label{fig:relicsub}
\end{figure}
\begin{figure}[h]!
    \centering
    \includegraphics[width=0.42\linewidth]{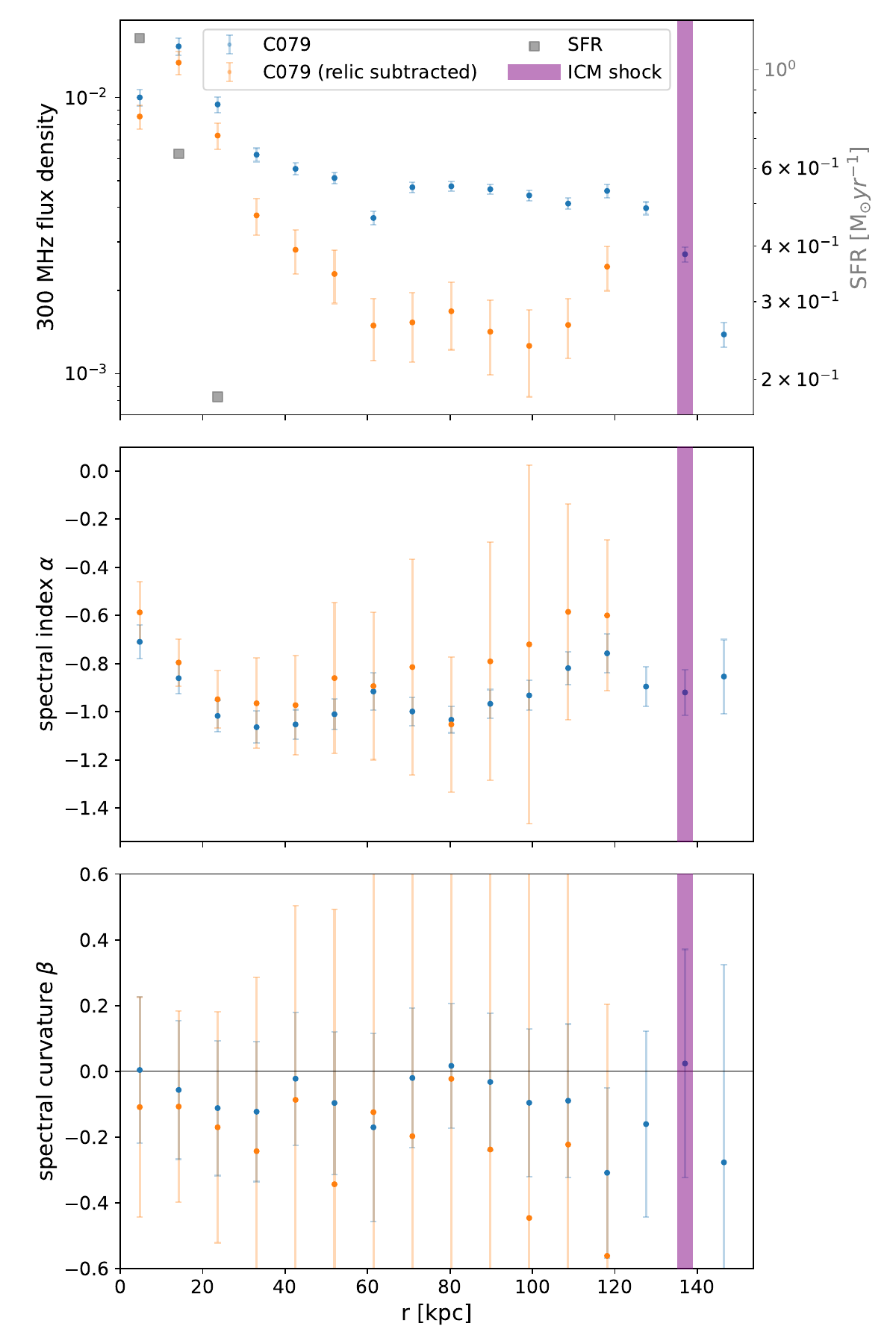}
    \includegraphics[width=0.42\linewidth]{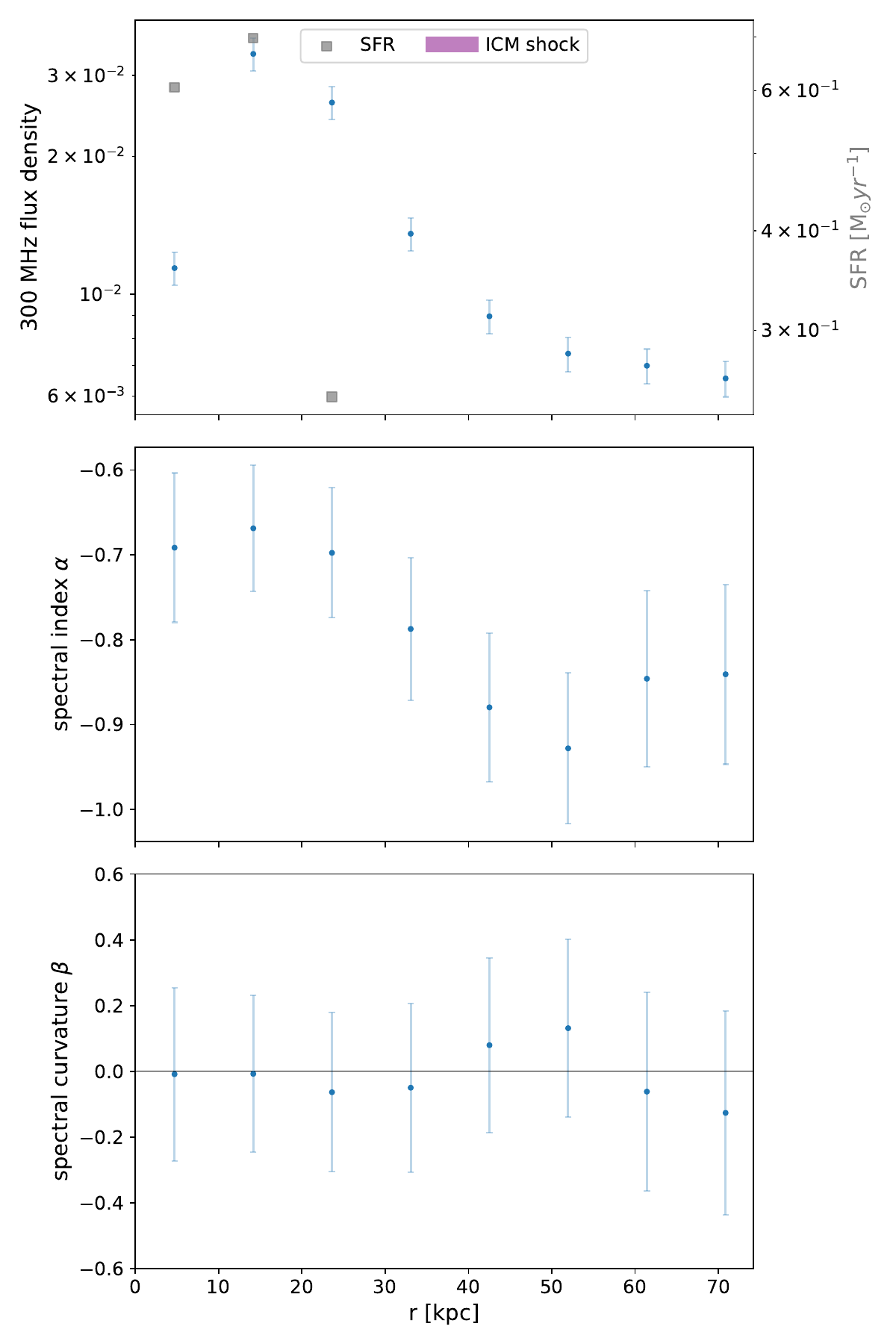}
    \caption{Flux density (top panel), SFR (top panel, right $y$ axis), spectral index (mid panel), and spectral curvature (bottom panel) as a function of projected distance along the C\,079 tail (left), and C\,073 tail (right). The vertical purple line shows the location of the ICM shock.}
    \label{fig:app_c079}
\end{figure}
\FloatBarrier
\newpage

\section{Color-color plots and uncertainty maps}
\label{sec:appendix_spec}
\begin{figure*}[h!]
    \centering
    \includegraphics[width=0.99\linewidth]{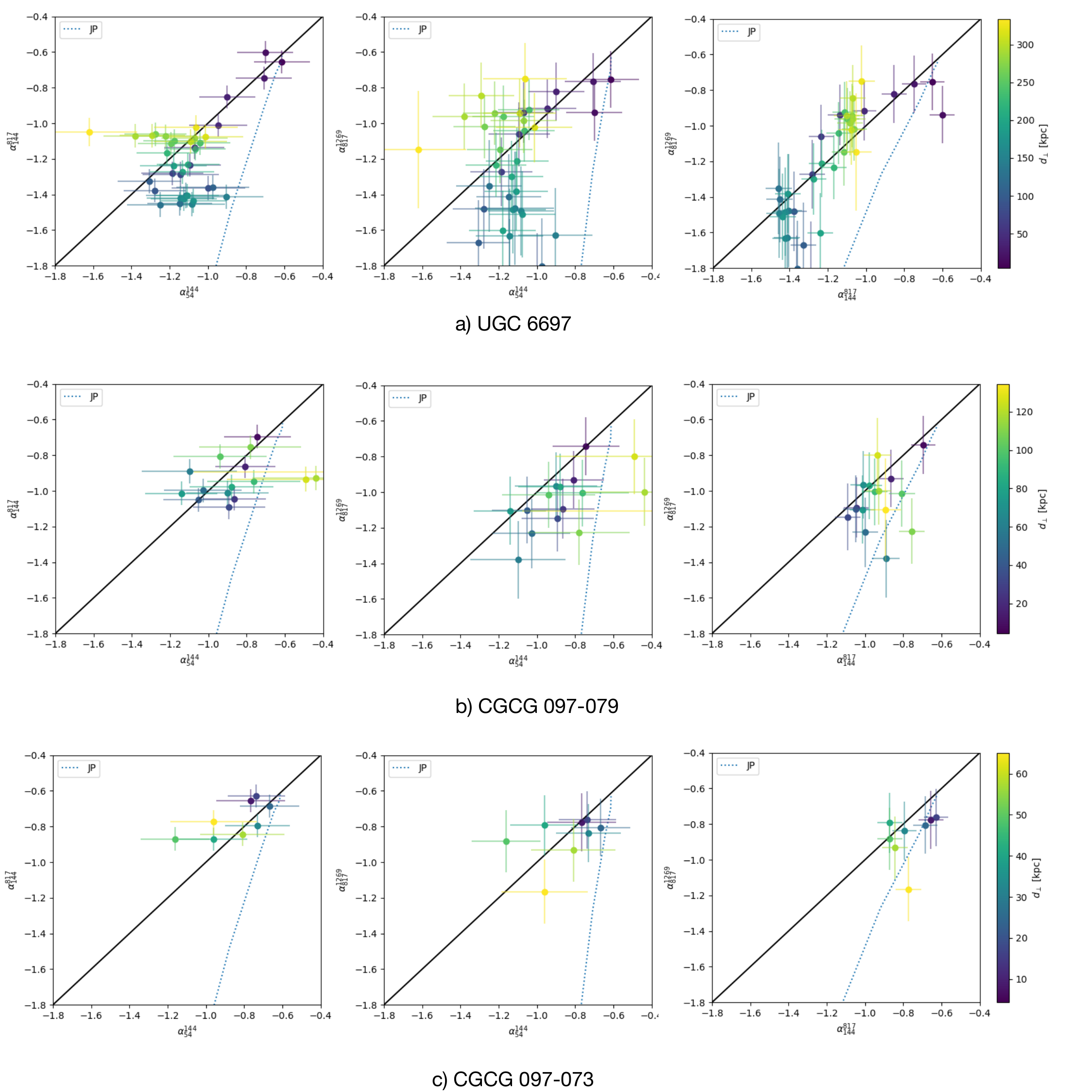}
    \caption{Color-color plots showing $\alpha_{54}^{144}$ vs. $\alpha_{144}^{817}$ (left column), $\alpha_{54}^{144}$ vs. $\alpha_{817}^{1269}$ (mid column), and $\alpha_{144}^{817}$ vs.  $\alpha_{817}^{1269}$ (right column) for the three galaxies (rows). The color of the points corresponds to the projected distance along the tails. The black line shows the power-law scenario and the dotted blue line the JP aging model.}
    \label{fig:color-color}
\end{figure*}
\begin{figure}
    \centering
    \includegraphics[width=0.45\linewidth]{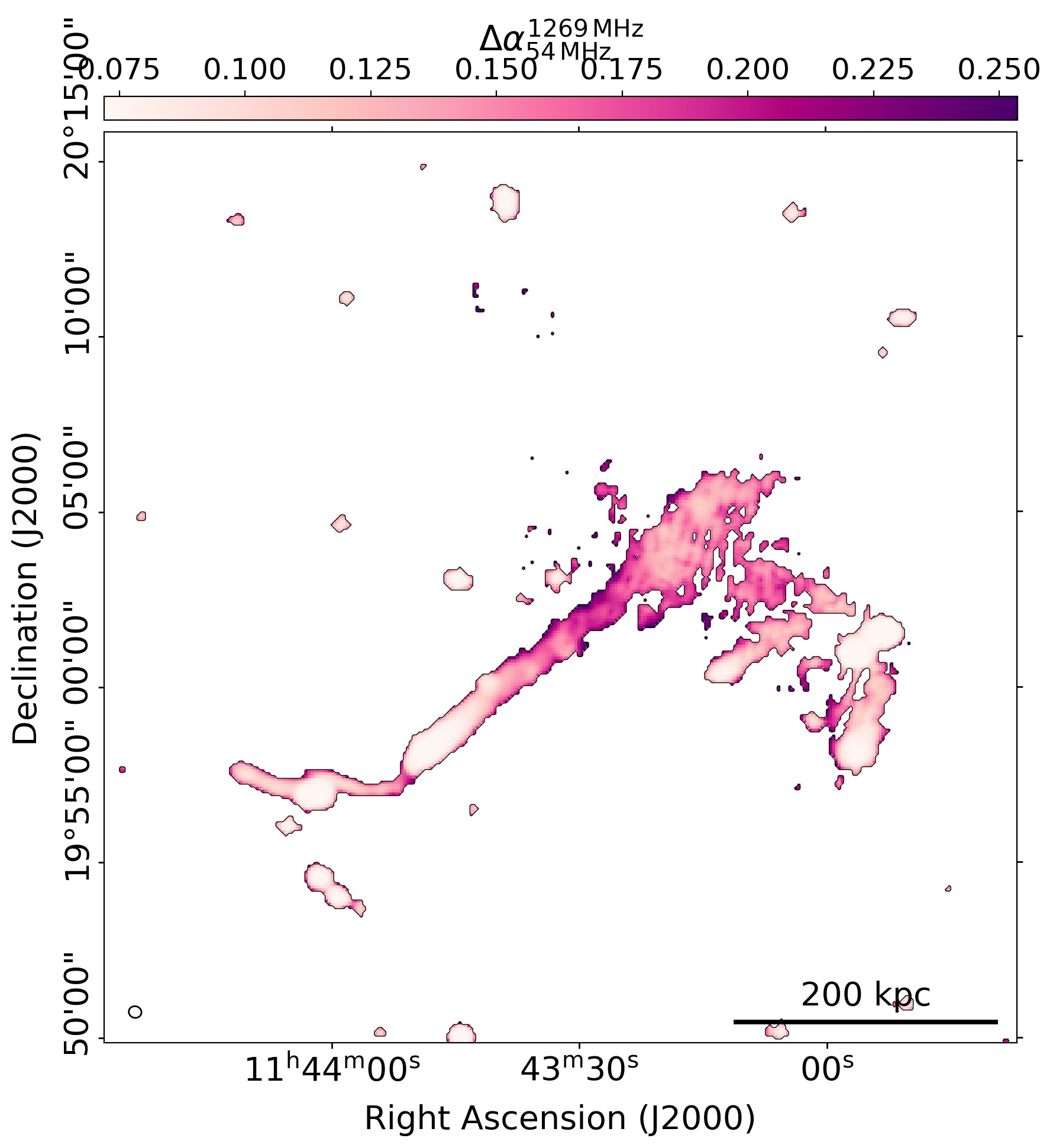}
    \includegraphics[width=0.45\linewidth]{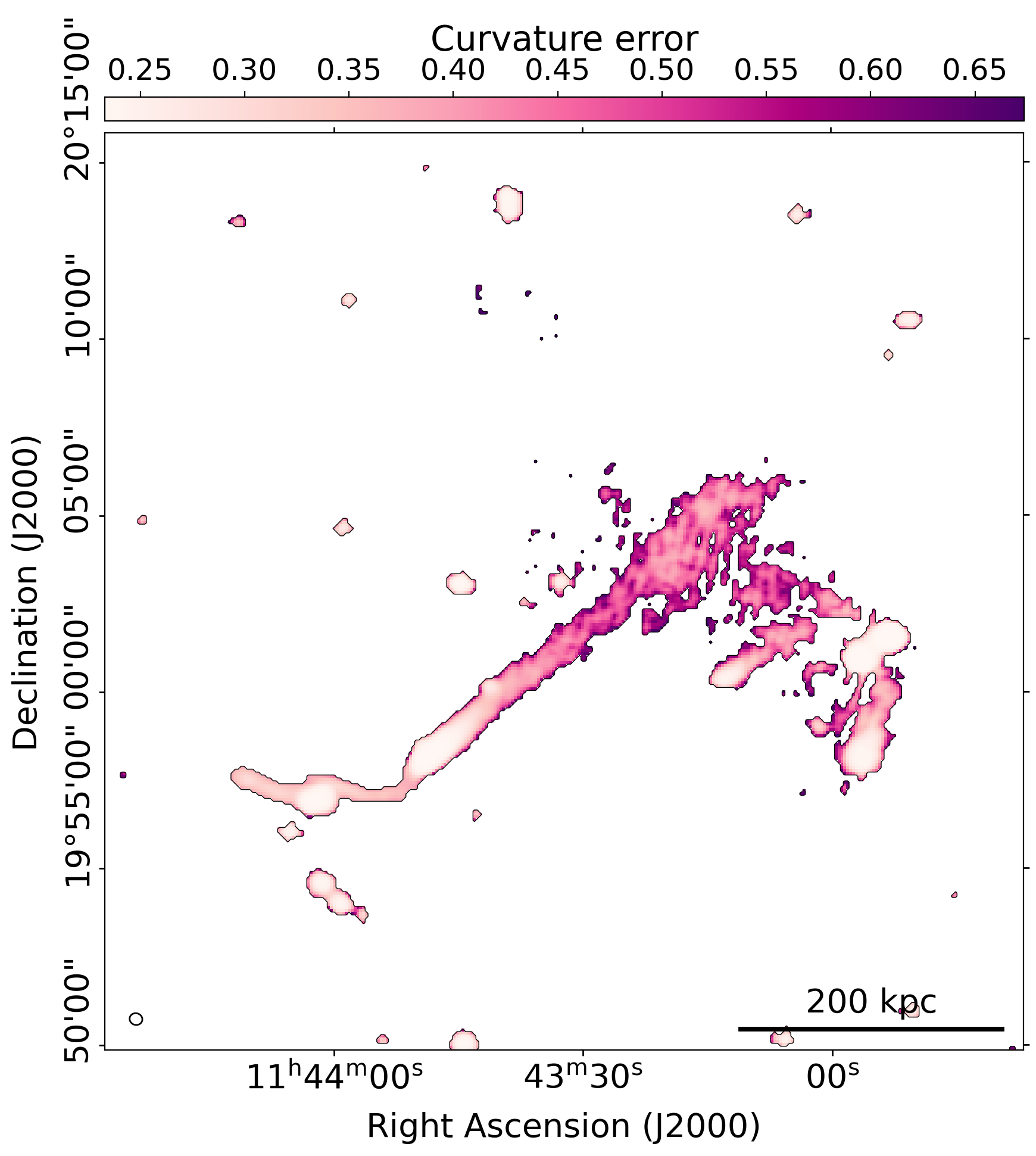}
    \caption{Uncertainty of the spectral index (top panel) and spectral curvature (bottom panel) maps presented in \autoref{fig:si}. Flux density scale uncertainties are taken into account.}
\end{figure}

\FloatBarrier
\section{Cosmic ray model}\label{sec:appendix_CRmodel}
\begin{equation}\label{eq:fullFP}
    \frac{\partial N(E,t)}{\partial t} = \frac{\partial}{\partial E} \left[ N(E,t) b(E)\right] - \frac{N(E,t)}{t_\mathrm{esc}} + Q(E,t).
\end{equation}
Our toy model consists of a series of closed-box models at different radii $r$. For each individual step, equation \autoref{eq:fokkerplanck}, a simplified form of the Fokker-Planck equation \autoref{eq:fullFP}, is evolved for the time $t_\mathrm{galax}$ since the galaxy passed. In practice, we solve the equation numerically employing the method of \citet{chang1970practical} as implemented in the \texttt{pychangcooper} library\footnote{\url{github.com/grburgess/pychangcooper}}. The energy grid is logarithmically spaced between $1\,$MeV and $100\,$GeV, and time steps of 10\,kyr were used. We tested our model on two cases: A) We verified that the maximum lifetime for synchrotron emission is indeed obtained with the minimum aging magnetic field $B_\mathrm{min}\approx1.9\,\mathrm{\mu{G}}$. B) We compared the aging-only spectrum after $100\,Myr$ to the ones obtained with identical parameters from the codes \texttt{synchrofit} and \texttt{BRATS \citep{Harwood2013}}. We found less than 10\% deviation between the different codes.
 
\end{appendix}
\end{document}